\documentclass[final,onecolumn,usenatbib]{mn2e}
\usepackage{graphicx,subfigure}
\usepackage{bm}
\usepackage{longtable}
\usepackage{amsmath,amssymb,graphicx}
\usepackage{rotating}
\usepackage{color}
\usepackage[normalem]{ulem}
\newcommand\nat{Nature}%
\newcommand\apj{ApJ}%
\newcommand\aap{A\&A}%
\newcommand\apjl{ApJL}%
\newcommand\mnras{MNRAS}%
\newcommand\prd{PhRvD}%
\newcommand{\be}{\begin{equation}}
\newcommand{\ee}{\end{equation}}
\newcommand{\bea}{\begin{eqnarray}}
\newcommand{\eea}{\end{eqnarray}}
\newcommand{\Lb}{\left(}
\newcommand{\Rb}{\right)}
\newcommand{\LB}{\left[}
\newcommand{\RB}{\right]}

\graphicspath{{./}{Figures/}}
\title[Prompt Emission Variability]{Gamma Ray Burst Prompt Emission Variability in Synchrotron and Synchrotron Self-Compton Lightcurves}
\author[Resmi \& Zhang]{Lekshmi Resmi$^{1,2}$, Bing Zhang$^1$\\
1. Dept. of Physics \& Astronomy, University of Nevada, Las Vegas, NV 89154-4002, USA.\\
2. Dept. of Astronomy \& Astrophysics, Tata Institute of Fundamental Research, Mumbai 400005, India.}
\begin{document}

\date{Accepted.....; Received .....}

\pagerange{\pageref{firstpage}--\pageref{lastpage}} \pubyear{}

\maketitle

\label{firstpage}

\begin{abstract}
{Gamma Ray Burst prompt emission is believed to originate from electrons accelerated in a highly relativistic outflow. ``Internal shocks'' due to collisions between shells ejected by the central engine is a leading candidate for electron acceleration. While synchrotron radiation is generally invoked to interpret prompt gamma-ray emission within the internal shock model, synchrotron self-Compton (SSC) is also considered as a possible candidate of radiation mechanism. In this case, one would expect a synchrotron emission component at low energies, and the naked-eye GRB 080319B has been considered as such an example. In the view that the gamma-ray lightcurve of GRB 080319B is much more variable than its optical counterpart, in this paper we study the relative variability between the synchrotron and SSC components. We develop a ``top-down'' formalism by using observed quantities to infer physical parameters, and subsequently to study the temporal structure of synchrotron and SSC components of a GRB. We complement the formalism with a ``bottom-up'' approach where the synchrotron and SSC lightcurves are calculated through a Monte-Carlo simulations of the internal shock model. Both approaches lead to the same conclusion. Small variations in the synchrotron lightcurve can be only moderately amplified in the SSC lightcurve. The SSC model therefore cannot adequately interpret the gamma-ray emission properties of GRB 080319B.
}
\end{abstract}



\section{Introduction}
\label{Intro}
Gamma Ray Burst (GRB) prompt emission lightcurves are complex with superimposed rapid short-scale variabilities. Variabilities of the order of milliseconds in the prompt phase, detected in $\gamma$-rays, were known since the discovery of the earliest GRBs and has led to the subsequent proposal of the ``internal shock'' model, where the energy in the relativistic flow is dissipated through multiple collisions within the ejecta. 

In the internal shock model \citep{1994ApJ...430L..93R}, the ultra-relativistic outflow from the central engine (ejecta) consists of a succession of shells with random lorentz factors. When a fast moving shell (with lorentz factor $\Gamma_f$) collides with one moving slowly (with $\Gamma_s$) ahead of it, a pair of internal shocks develops which dissipates the kinetic energy in the flow. Each pulse in the burst lightcurve corresponds to one such collision \citep{1997ApJ...490...92K, 2009ApJ...707.1623M}. The physical parameters of the dissipation region, magnetic field and electron distribution, depend on the masses and initial lorentz factors of the colliding shells, and the unknown microphysics of relativistic shocks \citep{1998MNRAS.296..275D}. Hence they vary erratically between collisions and so does the final flux.

Almost in all cases, prompt emission has been observed only in the narrow $\gamma$-ray band until very recently. This has limited our understanding of the underlying emission process. Observed spectra suggest that the radiative process is non-thermal. The most likely candidate is synchrotron radiation. Nonetheless, synchrotron self-Compton (SSC) process has been also suggested (e.g. \citep{2000ApJ...544L..17P, 2008MNRAS.384...33K}). For such models, one would expect a synchrotron component peaking in the lower energy band, and prompt optical emission is expected. In recent years, a few rapidly responding GRB-dedicated optical telescopes (e.g. RAPTOR, TORTORA, ROTSE) have become instrumental in detecting optical emission simultaneous to the $\gamma$-ray burst \citep{2005Natur.435..178V, 2006Natur.442..172V}. Most often these detections were limited to a few observations in the entire duration of the burst. Nevertheless, in a few cases a temporal correlation could be established between the optical and the $\gamma$-ray lightcurves, indicative of their possible origin from the same dynamical process \citep{2007ApJ...657..925Y,2007ApJ...663.1125P}. This improved spectral coverage has led to a better understanding of the prompt emission region \citep{2009MNRAS.398.1936S}. 

The optical flash of GRB080319B seen in unison with the $\gamma$-ray emission was exceptionally bright \citep{2008Natur.455..183R}. Optical prompt emission was observed throughout the entire duration of the $\gamma$-ray component with remarkable time resolution. The onset is simultaneous in both bands and the overall shape of the lightcurves are similar, indicating that emission in the two bands are possibly physically related. Flux in V-band was almost four orders of magnitude higher than the extrapolation of the $\gamma$-ray spectrum, implying that the two lightcurves are likely to have originated from two different emission processes. 

High and low energy emission tracking each other but belonging to two different radiative processes naturally led to the conjecture that optical prompt emission in GRB080319B is due to synchrotron mechanism in the internal shocks and these photons were up-scattered to the $\gamma$-ray band by the SSC process \citep{2008Natur.455..183R,2008MNRAS.391L..19K}. Despite its advantage of interpreting the rough tracking behavior between the two bands, this model also has several difficulties. For example, a few seconds lag between the two lightcurves is not straightforwardly expected in this model. Several later calculations claimed that the emission radius required under this scenario will be much larger if internal shocks were to occur \citep{2009MNRAS.395..472K,2009ApJ...692L..92Z}. Another drawback of this model is the energy crisis that occurs due to the presence of the bright 2nd order SSC component \citep{2009MNRAS.393.1107P,2009A&A...498..677B,2009ApJ...692L..92Z}. The non-detection of this second SSC bump in the prompt emission spectra as observed by Fermi LAT (e.g. \cite{2009ApJ...706L.138A,2011ApJ...730..141Z}) also places a great constraint on the synchrotron + SSC model.
Alternative models to interpret the rough tracking optical/$\gamma$-ray behavior of GRB 080319B have been proposed. \cite{2009PhRvD..79b1301F} advanced the idea of a neutron loaded fireball where both optical and $\gamma$-ray emission are synchrotron in origin but from two different electron populations, one being the original electrons in the plasma while the other originates from the $\beta$-decay process. 
\cite{2009ApJ...692.1662Y} suggested that a pair of internal forward and reverse shocks could be responsible for the $\gamma$-ray and optical emission respectively. 
Acknowledging the difficulty of the simplest internal shock SSC model, \cite{2009MNRAS.395..472K} invoked relativistic turbulence to improve the SSC model (cf. \cite{2009ApJ...695L..10L}).

One interesting observational feature of GRB 080319B is that its $\gamma$-ray lightcurve is much more variable than its optical counterpart \citep{2008Natur.455..183R}. The time resolution of optical observation is poorer than $\gamma$-rays, but even if one re-bins the $\gamma$-ray lightcurve to the same temporal resolution as the optical lightcurve, the $\gamma$-ray lightcurve still appears much more variable. This feature would give important constraints on the models. For example, the FS/RS internal shock model \citep{2009ApJ...692.1662Y} would predict a similar variability in both the optical and $\gamma$-ray lightcurves, so it cannot interpret the above feature. The two-zone model \citep{2009PhRvD..79b1301F}, on the other hand, is more consistent, since the optical emission is expected to occur at a larger radius, where the angular spreading time is longer. The synchrotron + SSC model \citep{2008MNRAS.391L..19K,2008Natur.455..183R} is more difficult to access since the relative variability between the two emission components has not been studied in the past.

In this paper, we study the relative variability within the framework of the synchrotron + SSC model.
We approach the problem through two complementary methods. In the `top-down' method, we use the observed optical lightcurve as the input synchrotron component, derive the fluctuations in the underlying physical parameters, and self consistently calculate the SSC lightcurve. In the `bottom-up' method, we follow the standard formalism to simulate the lightcurves in the frame work of the internal shock model. We generate a set of basic physical parameters through Monte-Carlo simulations, calculate both the synchrotron and SSC lightcurves and compare the fluctuations. The aim is to compare the relative variability between the two lightcurves and then address whether the observational features of GRB 080319B can be interpreted.

In section-2 and section-3 respectively, we describe our methods and results from the two approaches mentioned above.
%
%
%
\section{Lightcurve calculation in the `top-down' method}
We first construct a `top-down' method where physical parameters are expressed in terms of the observed optical luminosity. This approach enables us to reconstruct temporal fluctuations of the physical parameters from the structure of the observed synchrotron lightcurve \citep{2010ApJ...719L..10B} and use it to estimate the corresponding variability that would appear in the SSC component. 

The synchrotron component is the input V-band lightcurve itself. We have to consistently estimate the SSC component that would have arisen from the optical photons and the electrons that have produced them. This would require knowledge of the bulk lorentz factor $\Gamma$ of the outflow, the distance $R$ of the emission region from the center of explosion, the co-moving magnetic field ($B^{\prime}$) in the dissipation region, the electron distribution $N({\gamma_e})$, and the ratio $\mathcal{Y}$ of SSC to synchrotron luminosity. The temporal structure of the observed optical lightcurve is the combined effect of the time evolutions of all these parameters. It is difficult but possible to disentangle each of these parameters from the optical lightcurve alone with some simplifications and assumptions.
%
%
\subsection{Magnetic field in the internal shock region}
In the internal shock scenario, magnetic fields are generated by the shocks in the dissipation region (e.g., \cite{1999ApJ...526..697M}). Since this process is poorly understood from theoretical considerations, empirical methods are followed where the magnetic energy density is assumed to be proportional to the dissipated thermal energy measured in the co-moving frame. If $L_w$ is the luminosity of the wind from the central engine, $\delta t$  is the typical variability time-scale and $\eta$ is the efficiency of energy dissipation, internal energy in the co-moving frame can be expressed as $(\eta L_w \delta t )/\Gamma$, where $\Gamma$ is the bulk Lorentz factor of the final shell after collision, which enters the expression through frame transformation. Assuming that the shells are of equal mass, $\Gamma$ can be written as $\Gamma_s \sqrt{a_g}$, where $a_g = \Gamma_f/\Gamma_s$. If the dissipation region is at a distance $R$ from the central engine, the co-moving volume can be written as $4 \pi R^2 \Delta R^\prime$, where $\Delta R^\prime$ is the co-moving width of the shell. This width can be approximated as $\Delta R^{\prime} = R/\Gamma$, and the variability time scale can be expressed as $\delta t=\frac{R}{2 \Gamma_{\rm{s}}^2 c} \frac{a_g^2 -1}{a_g^2}$. This leads to the final expression of the co-moving magnetic energy density $u^{\prime}_B = {B^{\prime}}^2/8 \pi = \epsilon_B \frac{1}{8 \pi} \eta \frac{a_g^2 -1}{a_g^2} \frac{L_w}{\Gamma_{s}^2 R^2 c}$ \citep{2002ApJ...581.1236Z}.

From theoretical considerations, the bolometric luminosity is essentially related to the luminosity of the outflow (wind) from the central engine, $L_w$. An amount $\eta L_w$ of the original wind luminosity is dissipated as internal energy via internal shocks, which is carried mostly by protons. Depending on the interaction between the protons and electrons in the plasma, a fraction $\epsilon_e$ of this thermal energy is transferred to the random kinetic energy of the electron pool, of which a fraction $\kappa$ is radiated away. In the fast cooling regime, it is valid to assume that all the kinetic energy available to the electron pool is converted to radiation ($\kappa = 1)$. Hence the bolometric luminosity can be written as
\be
L_{\rm{bol}} = \epsilon_e \eta L_w
\label{eqA}
\ee

Using equation-\ref{eqA} to replace $\eta L_w$ in the expression of $u^{\prime}_B$, one can express the co-moving magnetic field strength $B^{\prime}$ as
\be
B^{\prime} \simeq 193 \sqrt{\frac{\epsilon_B}{\epsilon_e} L_{\rm{bol},52}} \sqrt{\frac{a_g^2-1}{a_g}} \frac{1}{R_{16} \Gamma_{300}} 
\label{eqmag}
\ee
where $R_{16}$ is $R$ in units of $10^{16}$~cm, $\Gamma_{300}$ is $\Gamma/300$ and $L_{\rm{bol,52}}$ is the bolometric luminosity in units of $10^{52}$~erg/sec.

We require to know the bolometric luminosity $L_{\rm{bol}}$ to calculate $B^{\prime}$. In the next section we describe how $L_{\rm{bol},52}$ can be written in terms of the observed optical specific luminosity $L_{\nu_V}$ and other physical parameters.
%
%
\subsection{From observed optical specific luminosity to bolometric luminosity}
The bolometric luminosity includes radiation emitted via both synchrotron and synchrotron self-Compton processes. Luminosity of the first order IC component can be expressed as, $L_{\rm{IC,1}} = \mathcal{Y}_1 L_{\rm{syn}}$, and that of the second order IC can be written as $L_{\rm{IC,2}}= \mathcal{Y}_1 \mathcal{Y}_2 L_{\rm{syn}}$ \citep{1996ApJ...473..204S,2007ApJ...655..391K}, where $\mathcal{Y}_1$ is the Compton parameter for the first order IC scattering defined by $L_{\rm{IC,1}}/L_{\rm{syn}}$, and $\mathcal{Y}_2$ is the Compton-parameter for the second order IC scattering defined by $L_{\rm{IC,2}}/L_{\rm{IC,1}}$.

Compton scattering between electrons of Lorentz factor $\gamma$ and synchrotron photons of frequency $\nu_{\rm{syn}}^{\prime}$ (measured in the co-moving frame of the relativistic ejecta) can be treated in the Thomson regime if $\gamma \frac{h \nu_{\rm{syn}}^{\prime}}{m_ec^2} < 1$. Using the characteristic frequency, $\nu=\frac{e}{2 \pi m_e c} \, B^{\prime} \gamma^2$, of a synchrotron photon emitted by an electron of Lorentz factor $\gamma$, this threshold leads to a limiting Lorentz factor $\gamma_{\rm{KN}}$ $\sim 3500 (B^{\prime}/1000 G)^{-1/3}$, above which Klein-Nishina corrections to the scattering cross-section become important. Alternatively, any SSC photon above a limiting frequency $\nu^{\prime}_{\rm{KN}}$, defined as $\gamma_{\rm{KN}} m_e c^2$, has undergone the scattering process that took place in the KN regime. The Klein-Nishina limiting frequency for the first order SSC is $\sim 300$~GeV $(B^{\prime}/1000 G)^{-1/3} \Gamma_{300}$, hence the SSC scattering leading to the soft-$\gamma$-ray emission can safely be assumed to be in the Thomson regime in the rest-frame of the electrons. As a result, one has $\mathcal{Y}_1 = \mathcal{Y}_{\rm{Th}}$. However, following the same argument, $\gamma_{\rm{KN},2}$ of the second order SSC scattering between electrons and the first order SSC photons is $117.2 (B^{\prime}/1000 G)^{-1/5} $ and $\nu_{\rm{KN,2}}$ is $\sim 18$~GeV $(B^{\prime}/1000 G)^{-1/5} \Gamma_{300}$. $\nu_{KN,2}$ could fall below $\nu_m^{ic2}$ of the 2nd order IC component, which means a fraction of the scattering events will require KN correction. As a result, the KN correction could very well be applicable for GRB080319B, and one is likely to have $\mathcal{Y}_2 = \mathcal{Y}_{\rm{KN}} < \mathcal{Y}_{\rm{Th}}$. %
Hence, the bolometric luminosity can be written as
\be
L_{\rm{bol}}(t) = (1+\mathcal{Y}_{\rm{Th}} + \mathcal{Y}_{\rm{Th}} \mathcal{Y}_{\rm{KN}})L_{\rm{syn}}(t)
\label{eqlbol}
\ee

Synchrotron luminosity $L_{\rm{syn}}$ can be estimated from the observed V-band specific luminosity $L_{\rm V}$, once we know the spectral regime the optical band belongs to. A precise estimation of this requires observation of the spectral index and break frequencies, which we do not have for GRB080319B. Nevertheless, since we are attempting a general framework for comparing synchrotron and SSC variability starting from the synchrotron lightcurve, we investigate all the possible spectral regimes.

The typical inferred magnetic field strength implies that in the prompt emission region, even the lowest energy (corresponding to a Lorentz factor $\gamma_m$) electrons of the injected spectrum are undergoing heavy radiative losses. Moreover, in order to keep energy requirements reasonable, electrons emitting in the optical band has to be radiatively efficient. Hence the optical V-band is expected to be in the `fast cooling' regime of the synchrotron spectrum. We check the consistency of this assumption using the value of $B^{\prime}$ obtained later in this section and find that this starting assumption is self-consistent. It is also assumed that the optical band is not self-absorbed. At the end of the procedure we check the consistency of this assumption as well. There are only two possible spectral regimes satisfying this condition: (i) $\nu_c < \nu_V < \nu_m$ and (ii) $\nu_c < \nu_m < \nu_V$. For the former, the optical specific luminosity $L_{\rm V}$ can be written as $L_{\nu_m} \Lb \frac{\nu_V}{\nu_m} \Rb^{-1/2}$ and for the latter it is $L_{\nu_m} \Lb \frac{\nu_V}{\nu_m} \Rb^{-p/2}$, where $L_{\nu_m}$ is the specific luminosity at $\nu_m$. For fast cooling electrons, the total synchrotron luminosity $L_{\rm{syn}}$ can be approximated as $L_{\nu_m} \nu_m$. Hence in terms of ${\nu_V}$, $L_{\rm{syn}}$ can be written as $L_{V} \sqrt{\nu_V \nu_m}$ for $\nu_c < \nu_V < \nu_m$ and $L_{V} \nu_m \LB \frac{\nu_V}{\nu_m}\RB^{p/2}$ for $\nu_c < \nu_m <\nu_V$.

Hence the bolometric luminosity in eq-\ref{eqlbol} is represented in terms of the two $\mathcal{Y}$-parameters ($\mathcal{Y}_{\rm{Th}}$ and $\mathcal{Y}_{\rm{KN}}$), $B^{\prime}$ and $\gamma_m$ (the latter two parameters entering the expression through $L_{\rm{syn}}$). 
\be
L_{\rm{bol}} = \left\{ 
\begin{array}{ll}
 (1+\mathcal{Y}_{\rm{Th}} + \mathcal{Y}_{\rm{Th}} \mathcal{Y}_{\rm{KN}}) L_{V} \sqrt{\nu_V \nu_m} &  {\mbox{for   }} \nu_c < \nu_V < \nu_m \\ 
 (1+\mathcal{Y}_{\rm{Th}} + \mathcal{Y}_{\rm{Th}} \mathcal{Y}_{\rm{KN}}) L_{V} \nu_m \LB \frac{\nu_V}{\nu_m}\RB^{p/2}  & {\mbox{for   }}   \nu_c < \nu_m <\nu_V
\end{array} \right.
\label{eqlbol2}
\ee

From eq-\ref{eqmag} and eq-\ref{eqlbol2} we can see that $L_{\rm{bol}}$ and $B^{\prime}$ depend on each other. We essentially require $B^{\prime}$ in the rest of the formalism. It can be obtained by algebraically solving eq-\ref{eqmag} and eq-\ref{eqlbol2}, and can be expressed in terms of other physical parameters. The final expressions of $B^{\prime}$ and $L_{\rm{bol}}$ are given in the appendix. The final expressions depend on the compton parameter $\mathcal{Y}$, which we will derive in the next section. $\gamma_m$ also appear in the final expressions; we will be using it as the input parameter (see section-2.4).
%
%

%
%
\subsection{Calculation of $\mathcal{Y}$ parameters} 
\label{sec23}

In the `top-down' method, The Compton-$\mathcal{Y}$ parameters enter the expression of the co-moving magnetic field through $L_{\rm{bol},52}$. They will be required in the calculation of the cooling frequency as well. Before deriving these parameters, we first introduce the relation between the compton-parameters and the ratio $\epsilon_e/\epsilon_B$ which is an important equality we use throughout the formalism. It is used in deriving the expressions in the appendix, and also in obtaining $\mathcal{Y}$ parameters in terms of $\gamma_m$.

\subsubsection{Relation with $\epsilon_e/\epsilon_B$}

For the 1st order scattering, the Compton $\mathcal{Y}$ parameter can be considered as the ratio of the SSC to synchrotron luminosity, which can be estimated as the ratio of the energy in seed photon field $U_{\rm{syn}}$ to that in the magnetic field $U_B$.  For the 2nd order scattering, it would be the ratio between the luminosities of the 2nd and 1st order IC components. $\mathcal{Y}$ is equivalent to $\kappa \frac{U_e}{U_B} \frac{1}{1+\mathcal{Y}+\mathcal{Y}^2}$ \citep{1996ApJ...473..204S, 2007ApJ...655..391K} in the limit where 1st and 2nd order SSC scattering are both in the Thomson regime. In the fast cooling regime, where $\kappa$ is nearly unity, this leads to the relation, $\epsilon_e/\epsilon_B =  \mathcal{Y} (1+\mathcal{Y}+\mathcal{Y}^2)$. However, if the 2nd order IC is in the KN regime (as is the case for our scenario in many runs), this expression is modified to
\be
\frac{\epsilon_e}{\epsilon_B} =  \mathcal{Y}_{\rm{Th}} (1+\mathcal{Y}_{\rm{Th}}+\mathcal{Y}_{\rm{Th}}\mathcal{Y}_{\rm{KN}})~.
\label{root}
\ee
\subsubsection{For the first order scattering : $\mathcal{Y}_{\rm{Th}}$}
We first estimate $\mathcal{Y}_{\rm{Th}}$, the Compton-$\mathcal{Y}$ parameter in the Thomson regime, valid for the first order IC scattering. For a fast cooling synchrotron spectrum, if $\nu_c < \nu_V < \nu_m$, the optical specific luminosity $L_{\rm V}$ can be expressed in terms of the peak luminosity $L_{\rm{max}}$ ($L_{\nu = \nu_c}$) as,
\be
L_{\rm V} = \sqrt{\nu_c/\nu_V} \, L_{\rm{max}}
\ee
where $L_{\rm{max}} = N_{\rm{rad}} \, \frac{\sqrt{3} e^3 B^{\prime}}{m_e c^2} \Gamma$ (Wijers \& Galama 1999). 

In terms of the total (both synchrotron and SSC) bolometric luminosity $L_{\rm{bol}}$, $L_{\rm V}$ can be expressed as
\be
L_{\rm V} = \frac{ L_{\rm{bol}} }{ (1+\mathcal{Y}_{\rm{Th}}+\mathcal{Y}_{\rm{Th}}\mathcal{Y}_{\rm{KN}}) } \, \frac{1}{\sqrt{\nu_V \nu_m}}
\ee
(here we have made use of eq-4).

After substituting for $L_{\rm{max}}$, $\nu_m$ (=$\frac{e}{2 \pi m_e c} B^{\prime} \gamma_m^2$), and $\nu_c$ (=$\frac{e}{2 \pi m_e c} B^{\prime} \gamma_c^2$), we obtain the relation
\be
\frac{L_{\rm{bol},52}}{{B^{\prime}}^2} = 1.95 \times 10^{-11} N_{\rm{rad},52} (1+\mathcal{Y}_{\rm{Th}}+\mathcal{Y}_{\rm{Th}}\mathcal{Y}_{\rm{KN}}) \gamma_m \gamma_c {\Gamma_{300}}^2.
\ee 

Since $\frac{L_{\rm{bol},52}}{{B^{\prime}}^2}$ can be substituted as $2.68 \times 10^{-5} (a_g/a_g^2-1) ({\epsilon_e/\epsilon_B}) R_{16}^2 \Gamma_{300}^2$ (using eq-2) and $(1+\mathcal{Y}_{\rm{Th}}+\mathcal{Y}_{\rm{Th}}\mathcal{Y}_{\rm{KN}})$ can be replaced by $(1/\mathcal{Y}_{\rm{Th}})(\epsilon_e/\epsilon_B)$ using eq-5, we arrive at, 
\be
\mathcal{Y}_{\rm{Th}} = 2.2 \times 10^{-6} \frac{a_g}{a_g^2-1} \frac{N_{\rm{rad},52}}{R_{16}^2} \gamma_m \gamma_c.
\label{eqtau}
\ee
Through the same approach, the above expression can be obtained for $\nu_m < \nu_V$ also.

It needs to be mentioned that, in the standard `bottom-up' approach, one starts from the above expression, derived by integrating the scattering cross-section over the electron energy spectrum assuming fast cooling electrons \citep{2000ApJ...543...66P, 2008MNRAS.384...33K}, substitutes for $\gamma_m$ in terms of $\epsilon_e$ and $\gamma_c$ in terms of $\epsilon_B$, and arrives at eq-5. 

\subsubsection{For the second order scattering : $\mathcal{Y}_{\rm{KN}}$}
For the second order scattering, which is likely to be in the Klein-Nishina regime, evaluation of the $\mathcal{Y}$-parameter is more complex. This is because unlike in the Thomson regime, the scattering cross-section depends on the Lorentz factor of each electron involved in the scattering. A full numerical calculation is required to obtain an exact estimate of $\mathcal{Y}_{\rm{KN}}$. In this paper, we adopt an analytical formalism where approximate estimates of the reduction due to Klein-Nishina effect in both scattering cross-section and the typical energy gain of the photons are used. Since in the SSC regime $\nu F_\nu$ peak is at $\nu_m^{\rm IC}$ for a fast cooling spectrum, the reduction in effective scattering cross-section can be scaled down as  
\be
{\cal R}_{\sigma} = \frac{\sigma_{\rm{KN}}(x_{\gamma_m})}{\sigma_T}
\ee
where $x_{\gamma_m} =  \gamma_m \frac{h \nu_m^{\rm{IC}}}{300 \Gamma_{300} \, m_e c^2}$ is the normalized energy of first order SSC photon in the restframe of the relativistic electron with lorentz factor $\gamma_m$. $\nu_m^{\rm{IC}}$ is calculated as $2 \gamma_m^2 \nu_m$. For $x_{\gamma_m} \gg 1$, one can approximate $\cal R_{\sigma}$ to be $\frac{3}{8x_{\gamma_m}}(\log{2x_{\gamma_m}}+\frac{1}{2})$.

$\mathcal{Y}_{\rm{KN}}$ can be written as $\mathcal{Y}_{\rm{Th}} {\cal R}_{\sigma} \Delta \mathcal{E}_{\mathcal{R}}$, where $\Delta \mathcal{E}_{\mathcal{R}}$ is the ratio between average energy gain for the photon in the second order scattering to that in the first order scattering. The average gain in the 2nd order scattering can be approximated as $\frac{\gamma_{KN,2} m_e c^2}{h {\nu_m^{ic,1}}^{\prime}}$. We divide it by the typical energy gain in the first order scattering $\gamma_m^2$ to obtain $\Delta \mathcal{E}_{\mathcal{R}}$. Using the expression $\gamma_{KN,2} = 117.2(B^{\prime}/1000 {\rm G})$ from section-2.3 and substituting for ${\nu_m^{ic,1}}^{\prime}$ as $2 \times 2.8 \times 10^{6} {\rm{Hz}} B \gamma_m^4$, we can reduce $\Delta \mathcal{E}_{\mathcal{R}}$ to be $\frac{4 \times 10^{13} \gamma_{\rm{KN},2}}{B \gamma_m^6}$.
Hence, when KN corrections are strong, ie., for $x_{\gamma_m} \gg 1$, the final expression for $\mathcal{Y}_{\rm{KN}}$ is
\be
\mathcal{Y}_{\rm{KN}} = \mathcal{Y}_{\rm{Th}} \, \frac{3}{8x_{\gamma_m}} \Lb \log{2x_{\gamma_m}}+\frac{1}{2} \Rb \, \frac{4 \times 10^{13} \gamma_{\rm{KN},2}}{B \gamma_m^6}.
\label{yknexp}
\ee

In our calculations we assume $\mathcal{Y_{\rm{KN}}}$ to be $\mathcal{Y_{\rm{Th}}}$ if $x_{\gamma_m} \leq 1$, else we use the above expression. It is not possible to analytically estimate the correction when the KN-effect is moderate ($x_{\gamma_m}$ is a few), hence we use the two asymptotic estimates.

Substituting for $\mathcal{Y}_{\rm{KN}}$, we can rewrite eq-5 in terms of $\mathcal{Y}_{\rm{Th}}$, $\epsilon_e/\epsilon_B$ and $\mathcal{R}_{\sigma} \Delta \mathcal{E}_{\mathcal{R}}$. It is detailed in the appendix. In figure-1 we present the behaviour of $\mathcal{Y}$ parameters.

If the 2nd order scattering is in the Thomson regime, both first and second order $\mathcal{Y}$ parameter will be the same and will depend only on the ratio $\epsilon_e/\epsilon_B$. However, if the 2nd order scattering is affected by KN-effects, the first order $\mathcal{Y}_{\rm{Th}}$ will depend on $\epsilon_e/\epsilon_B$ and the term $ \mathcal{R}_{\sigma} \Delta \mathcal{E}_\mathcal{R}$. This term signifies the extent of KN effect (For scattering in Thomson regime, it is unity. The more $x_{\gamma_m}$ is, the smaller will be $ \mathcal{R}_{\sigma} \Delta \mathcal{E}_{\mathcal{R}}$ ). We estimated $\mathcal{Y}_{\rm{Th}}$ for a range of $\epsilon_e/\epsilon_B$ and $\mathcal{R}_{\sigma} \Delta \mathcal{E}_\mathcal{R}$; we find that $\mathcal{Y}_{\rm{Th}}$ has a strong dependence on the value of $\epsilon_e/\epsilon_B$, and the dependence on $ \mathcal{R}_{\sigma} \Delta \mathcal{E}_\mathcal{R}$ is fairly weak. Only in cases of very high $\epsilon_e/\epsilon_B$ does the dependence on the 2nd term become important. In addition to that, it is easy to note from eq-8 that for a given value of $\epsilon_e/\epsilon_B$, a lower $\mathcal{Y}_{\rm{KN}}$ (which implies stronger KN effect and larger of $x(\gamma_m)$) will result in a higher $\mathcal{Y}_{\rm{Th}}$.

\begin{figure*}
\includegraphics[scale=0.2,angle=270]{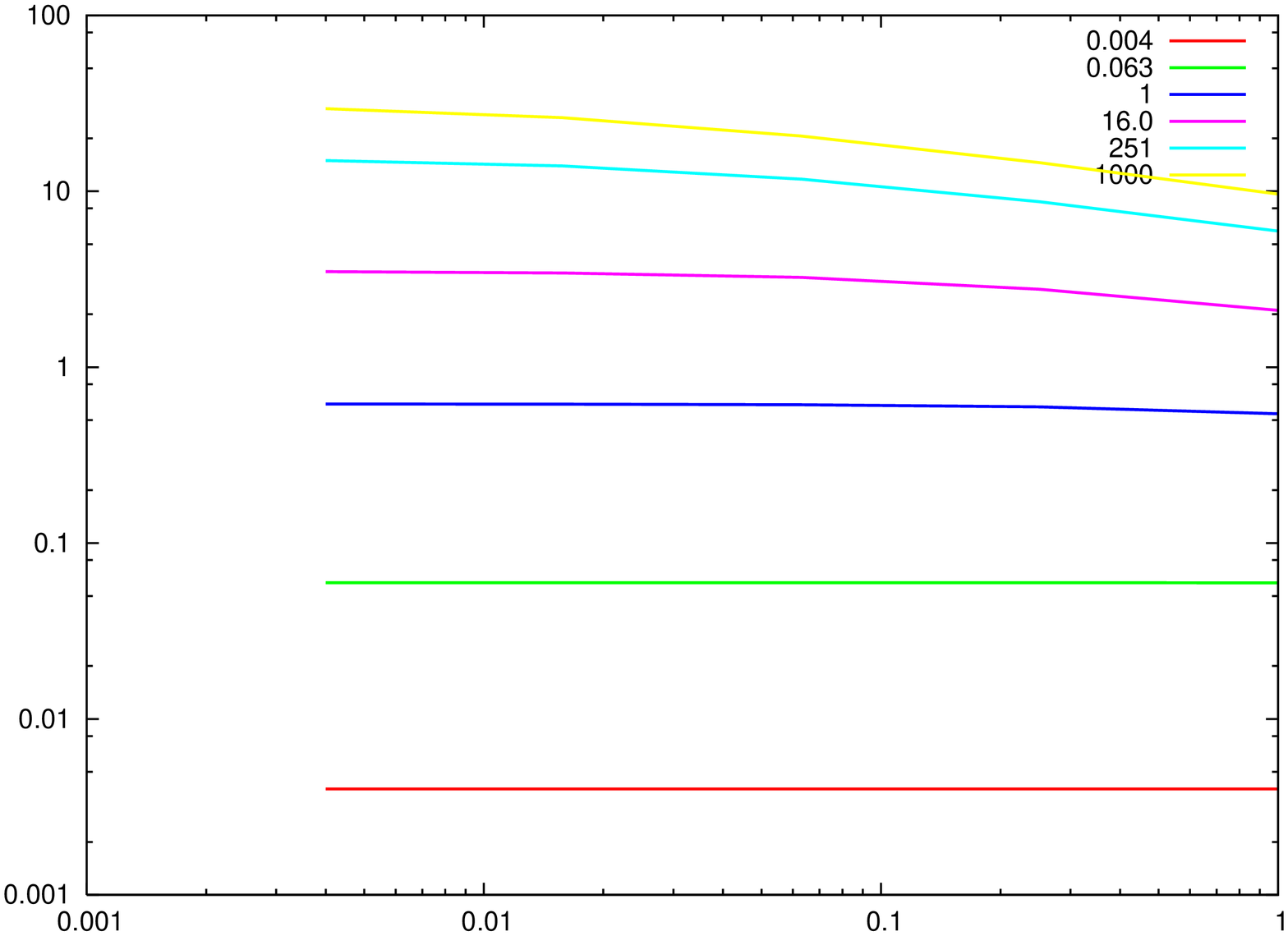}
\includegraphics[scale=0.2,angle=270]{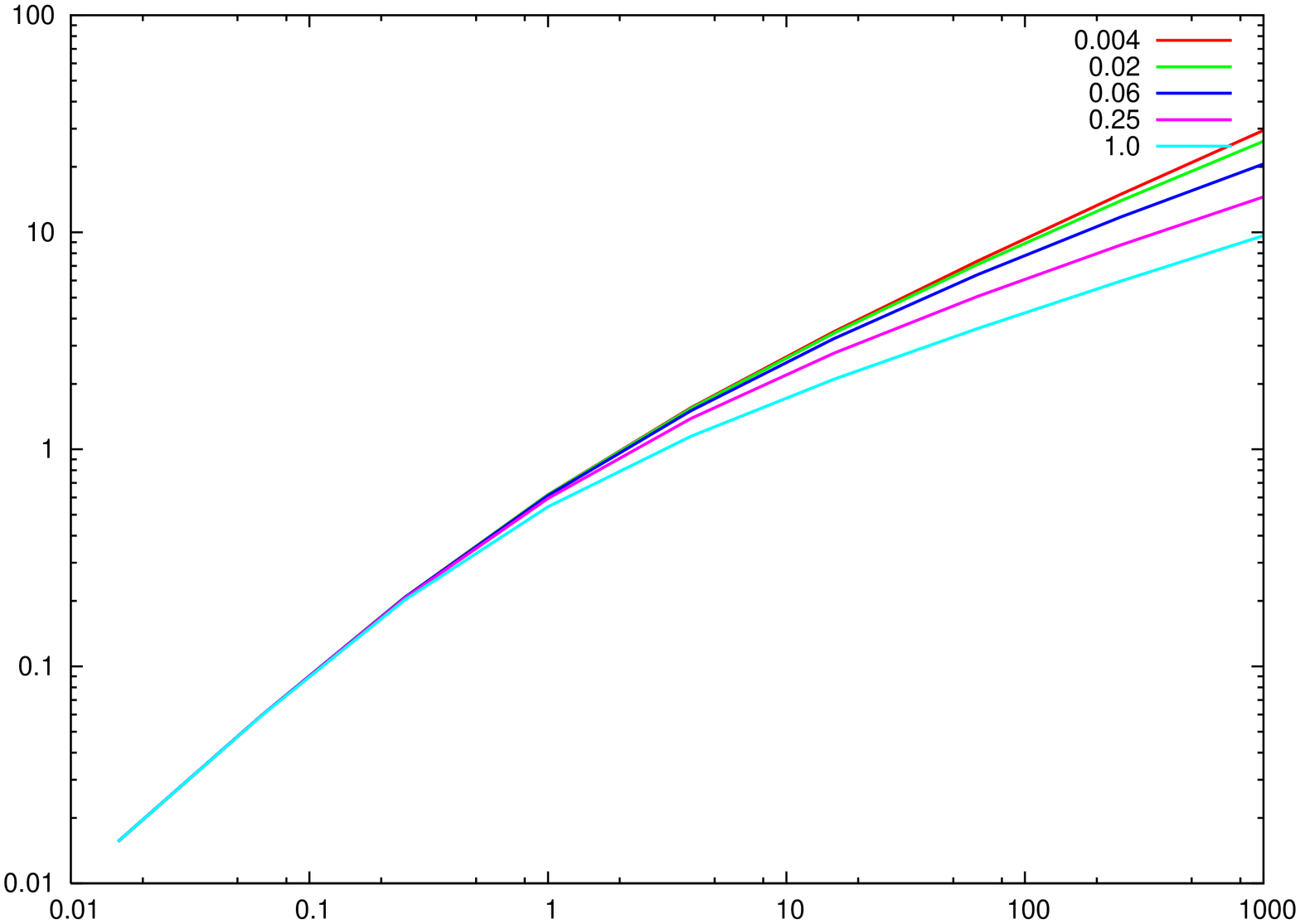}
\includegraphics[scale=0.2,angle=270]{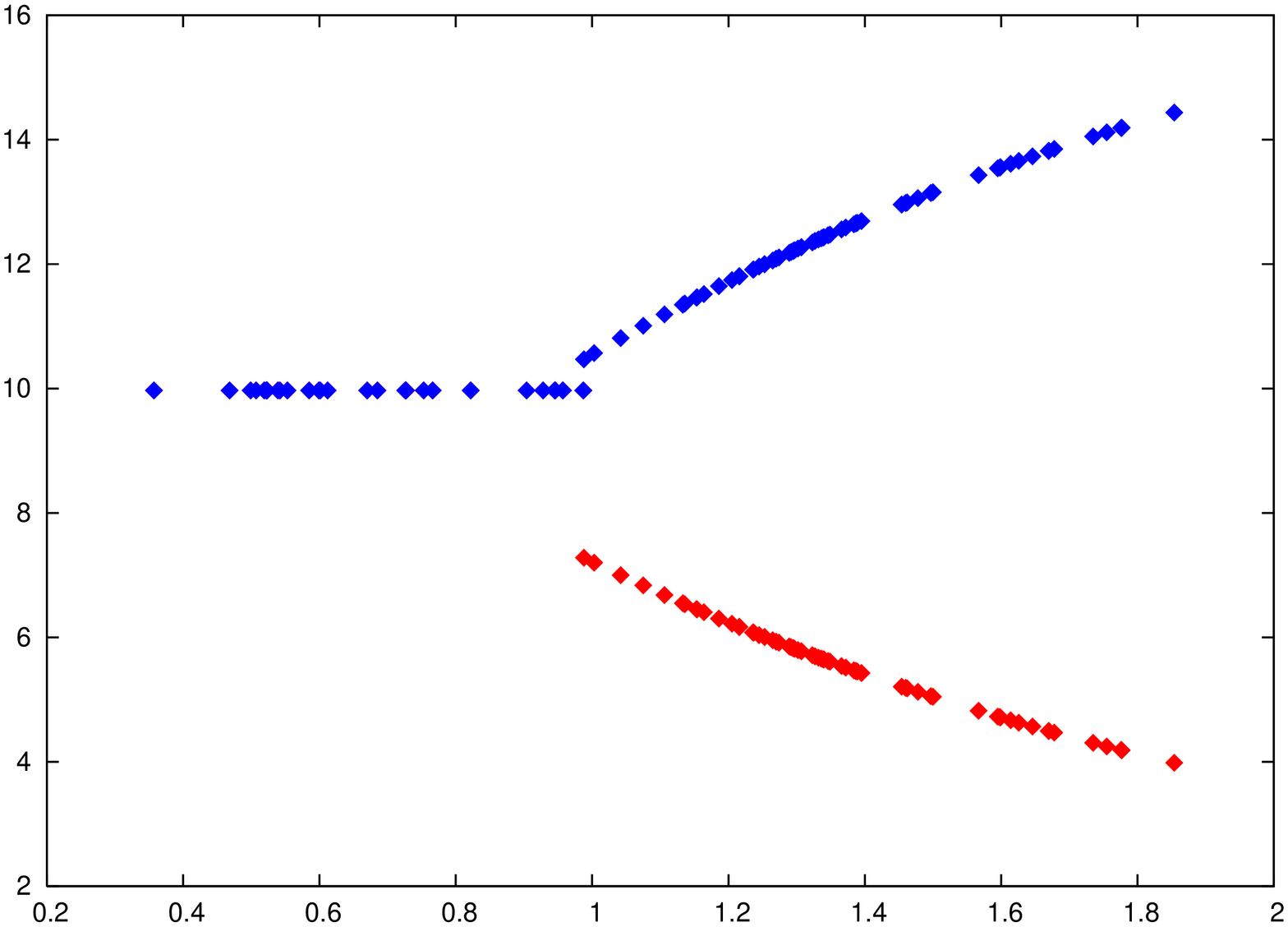}
\put(-374,-102){\scriptsize \bf {$\mathcal{R}_{\sigma} \Delta \mathcal{E}_\mathcal{R}$}}
\put(-445,-50){\scriptsize \bf {$\mathcal{Y}_{\rm{Th}}$}}
\put(-225,-102){\scriptsize \bf{$\epsilon_e/\epsilon_B$}}
\put(-295,-50){\scriptsize \bf {$\mathcal{Y}_{\rm{Th}}$}}
\put(-146,-70){\rotatebox{90}{$\mathcal{Y}$ parameters}}
\put(-75,-102){\scriptsize \bf{$x_{\gamma_m}$}}
\caption{The behaviour of $\mathcal{Y}_{\rm{Th}}$ (the $\mathcal{Y}$-parameter for the 1st order scattering), if the 2nd order SSC scattering is in the KN-regime. $\mathcal{Y}_{\rm{Th}}$ is obtained through non-linear root finding from eq-21. In the first figure, $\mathcal{Y}_{\rm{Th}}$ is plotted against $\mathcal{R}_{\sigma} \Delta \mathcal{E}_\mathcal{R}$ for various values of  $\epsilon_e/\epsilon_B$, in the middle figure,  $\mathcal{Y}_{\rm{Th}}$ is plotted against $\epsilon_e/\epsilon_B$ for a range of $ \mathcal{R}_{\sigma} \Delta \mathcal{E}_\mathcal{R}$. We can see that $\mathcal{Y}_{\rm{Th}}$ is highly sensitive to $\epsilon_e/\epsilon_B$ while the changes in $\mathcal{R}_{\sigma} \Delta \mathcal{E}_\mathcal{R}$ is not very consequential. In the 3rd panel, the two $\mathcal{Y}$ parameters, computed using eqn-21 in the appendix through a root-finding algorithm, are plotted for a range of $x(\gamma_m)$.  Here we have used an input $\epsilon_e/\epsilon_B$ of $1000$, $200 < \gamma_m < 230$, $\Gamma_{300}$ of $1.3$, $R_{16}$ of $0.3$ and the input optical lightcurve for $L_V$. For $x(\gamma_m) << 1$, $ \mathcal{Y}_{\rm{KN}} \sim \mathcal{Y}_{\rm{Th}}$, but as $x(\gamma_m)$ increases, the scattering enters to the KN-regime, the $\mathcal{Y}$ parameter for the 2nd order scattering (in red) decreases and the $\mathcal{Y}$ parameter for the 1st order scattering (in blue) increases.}
\end{figure*}

%
\subsection{Sequential steps to estimate SSC emission}
\label{sectseq}
To calculate the SSC component, we require the final expressions of $B^{\prime}$, $N_{\rm{rad}}$, $\gamma_m$, $\mathcal{Y}_{\rm{Th}}$ and $\mathcal{Y}_{\rm{KN}}$ (we call these {\textit{class-1}} parameters), in addition to the input parameters (we will note them as {\textit{class-2}}) $a_g$, $\delta t$, $\Gamma_s$, $p$ (required only for the spectral regime case-ii), and $\epsilon_e/\epsilon_B$. We can see from the above sections that the five quantities in {\textit{class-1}} are all inter connected. The key is to find a way to disentangle them, start from one and to arrive at all the remaining quantities. 

A random number generator is used to determine the value of $\Gamma_s$ for a given pulse. $R_{16}$ is computed using the relation of radius $R = 2 c \delta t \Gamma_s^2 a_g^2/(a_g^2 - 1)$. $\Gamma_{300}$ is calculated as $\sim \sqrt{a_g} \Gamma_s/300$. Across the burst, typical fractional variation in $\Gamma_s$, $\delta \Gamma_s/\Gamma_s \sim 1/20$ to $1/10$. In a given simulation run $a_g$ and $\delta t$ are kept constant, $\delta t$ is kept somewhere in a range of $0.5$ -- $2.0$, and $a_g$ is fixed around $2 - 5$. In a given run, their values are chosen such that the variation in $R_{16}$ for the burst could be up to a factor of $3$ to $4$, to reproduce the range of radii in which shocks occur \citep{2000A&A...358.1157D}. Between various realizations of the simulation, $\Gamma_s$ varies from $50$ to $500$. The resultant range of $\Gamma_{300}$ through multiple runs is $ \sim 1$ to $5$. Multiple simulations resulted in a large range of $R_{16}$, from $0.01$ to $\sim 1$. 

Since $\gamma_m$ is roughly constant for a given pulse during shock crossing, we use $\gamma_m$ to be the input parameter, and randomly assign its value for a given pulse. The range of $\gamma_m$ varied from $\sim 50$ to $\sim 500$ through the runs. Whenever the optical frequency fell below $\nu_a$, we moved to another set of parameters. Depending on the other parameters, $\epsilon_e/\epsilon_B$ was varied such that $\mathcal{Y}_{\rm{Th}}$ is around $10$.

Eq-\ref{root2} connects $\mathcal{Y}_{\rm{Th}}$ with $B^{\prime}$ and $\gamma_m$. On the other hand, $B^{\prime}$, as we can see from its detailed expression given in appendix, depends on $\gamma_m$ and $\mathcal{Y}_{\rm{Th}}$. Other than that, only the quantities in {\textit{class-1}} enter in these two expressions. Hence, we can use the expressions of $B^{\prime}$ in the appendix and write eq-\ref{root2} in terms of $\gamma_m$ alone. However, we are not presenting this long expression (in fact one each for the two spectral regimes, $\nu_V < \nu_m$ and $\nu_V > \nu_m$). Below are the sequential steps we follow in estimating the SSC emission.

{\underline{step-1:}} $\mathcal{Y}_{\rm{Th}}$ is obtained from eq-21 for a given $\gamma_m$ and {class-2} parameters using non-linear root finding algorithms. 

{\underline{step-2:}} Now we know both $\gamma_m$ and $\mathcal{Y}_{\rm{Th}}$, so $B^{\prime}$ can be calculated. 

{\underline{step-3:}} Knowing $B^{\prime}$, $\mathcal{Y}_{\rm{Th}}$ and $\gamma_m$, we can now calculate $\mathcal{Y}_{\rm{KN}}$ from the expression in the appendix. 

In the code, the possibility of the 2nd order scattering to be in the Thomson regime is checked by monitoring the value of $x_{\gamma_m}$ once $B^{\prime}$ is estimated, and in that case the quantities are re-estimated. 

{\underline{step-4:}} Once we know $\mathcal{Y}_{\rm{Th}}$, $\mathcal{Y}_{\rm{KN}}$ and $B^{\prime}$ we can calculate $\gamma_c$. If $\gamma_c$ falls below unity, we set it to unity. 

Since we know $\nu_m$ and $\nu_c$, we can now do a self-consistency check about the synchrotron spectral regime, and see whether our initial assumption of fast cooling and the location of the optical frequency is valid or not. If the consistency is violated, we redo the calculations with a different set of parameters (especially $\gamma_m$ values).

{\underline{step-5:}} Finally, we use eq-\ref{eqtau} to estimate the value of $N_{\rm{rad}}$ with $\gamma_m, \gamma_c$ and $\mathcal{Y}_{\rm{Th}}$. 

We thus have all the quantities in {\textit{class-1}} known, and can compute the SSC lightcurve. The characteristic frequencies are estimated as $\nu_c^{IC} = 2 \gamma_c^2 \nu_c$ and $\nu_m^{IC} = 2 \gamma_m^2 \nu_m$, and the spectral normalization at $\nu = \nu_c^{IC}$ is estimated as $L_{\nu, \rm{max}} \mathcal{Y}_{\rm{Th}}/(\gamma_m \gamma_c)$. We follow \cite{2007MNRAS.380...78G} in calculating the SSC spectrum. We assume the flux to decay following the curvature effect, by $(t/t_p)^{-2+\beta}$ ($\beta$ being the spectral index), after the peak ($t_p$) of each pulse \citep{1996ApJ...473..998F, 2000ApJ...541L..51K, 2006ApJ...642..354Z, 2006ApJ...646..351L}.
\subsection{Measure of Variability}
An important point to note is that for GRB080319B the time resolution of the BAT lightcurve is almost $\sim 20$ times higher than that of the V-band lightcurve from TORTORA \citep{2008Natur.455..183R}. Many individual pulses and fine features may have been lost in the coarse resolution of the optical lightcurve. Therefore a direct comparison between our synthesized SSC lightcurve and the BAT lightcurve is not entirely appropriate. We can only make a qualitative study on the extent of amplification possible by the SSC process. Our aim is to see how small-scale fluctuations of the synchrotron lightcurve would appear in the SSC component, and how variations in the electron distribution can control their appearance.

The template for our analysis is the optical V-band lightcurve of GRB080319B. However, since our aim is to see whether the high variability of its $\gamma$-ray lightcurve is due to enhancement of small scale synchrotron fluctuations by Compton upscattering, we synthesize miniscule fluctuations by adding random Gaussian fluctuations at mid-points after interpolating the lightcurve. We synthesize multiple lightcurves using Gaussian distributions with different widths. We first test our method in the original optical lightcurve from TORTORA and later use the new synthesized lightcurves. In figure-2, we present one of the template lightcurves with synthesized fluctuations along with the original lightcurve from TORTORA. There are a total of eight pulses in the lightcurve, of which three are too miniscule that they would not have been discerned as independent pulses typically. Observations exist during the tail only for the 3rd, 5th, 7th and 8th pulses. To obtain the beginning of the pulse $t_0$, we fit the decay-part of these pulses with the temporal profile expected from the curvature-effect, which requires the tail to fall as $(t-t_0)^{-(2+\beta)}$. For pulse-8, which has good sampling over the tail, we could constrain the values of both $t_0$ ($18.0 \pm 14.0$ and the index ($-2-\beta = -2.5 \pm 1.1$). For the other three pulses, we have assumed $\beta$ to be around $0.5$ and obtained the value of $t_0$. For those pulses for which we could not run a fitting routine, we used the same value of $t_0$ as its nearest pulse. It is to be noted that the optical spectral index can be much steeper, especially because of the large spectral index observed in $\gamma$-rays. 

$\gamma_m$ is the parameter that we tune to control the pulse amplification in SSC. For the above synchrotron lightcurve, there will be a set of eight $\gamma_m$ values as input. The spectral regimes where the optical and the gamma-ray bands fall play a crucial role in the nature of the output gamma-ray lightcurve. Given the same distribution of $\Gamma_{300}$ and $R_{16}$, we see that if $\nu_{\rm{V}}$ is between $\nu_c$ and $\nu_m$ and if $\nu_\gamma$ is between $\nu_c^{\rm{IC}}$ and $\nu_m^{\rm{IC}}$, the two lightcurves become nearly identical (even with large variation within the input $\gamma_m$ set) because of the same functional dependence both synchrotron and SSC flux has on $\gamma_m$. This is true for the case where $\nu_m < \nu_{\rm{V}}$ and $\nu_m^{\rm{IC}} < \nu_\gamma$ as well. The best contrast is possible if $\nu_{\rm{V}} < \nu_m$, but $\nu_\gamma > \nu_m^{\rm{IC}}$ or $\nu_\gamma < \nu_c^{\rm{IC}}$. This is because of the additional dependence the SSC flux has on $\gamma_m$. 

  It is easy to see, using the expression for characteristic frequencies, that $\nu_{\rm{V}}$ will be below $\nu_m$ if $B^{\prime} \gamma_m^2 > 2 \times 10^{11}$. Similarly, $\nu_m^{\rm{IC}}$ will be below $650$~keV if $B^{\prime} \gamma_m^4 < 5 \times 10^6$. Since in our formalism $B^{\prime}$ depends on $\gamma_m$ and other input parameters (class-2 parameters), this basically boils down to a range of $\gamma_m$ for given class-2 parameters and the input synchrotron luminosity $L_{\rm V}$. But since the $\gamma_m$ dependence of $B^{\prime}$ comes through the non-linear expression in terms of $\mathcal{Y}_{\rm{Th}}$, we cannot give an analytical expression describing the range.

Amplification or quenching of the SSC pulse will depend on the input $\gamma_m$ of that pulse. An analytical expression for $L_p^{\rm{SSC}}$ -- $\gamma_m$ relation is difficult to obtain because of the non-linear nature of the algorithm. We observe the variation of $L_p^{\rm{SSC}}$ for a large range of $\gamma_m$ values and see that the pulse peak $L_p$ is roughly proportional to $\gamma_m^{3.3}$. Hence, the ratio between two adjacent peaks $L_p^i/L_p^{i+1} \propto (\gamma_m^i/\gamma_m^{i+1})^{3.3}$ if $\Gamma_{300}$ and $R_{16}$ do not vary much between pulses. Ideally, any contrast between adjoining pulses can be produced by changing $\gamma_m$, but the pulse profile will appear unnatural with sharp dips and rises for stark contrasts. The variability of SSC lightcurve depends on the distribution of $\gamma_m$ through pulses. In figure-3 we present a sample of $\gamma$-ray lightcurves we obtained along with the input optical synchrotron lightcurves. In the first panel, we have also given the output synchrotron lightcurve along with the input lightcurve (the only difference is the decay determined by the curvature-effect in case of the synthesized lightcurves) for comparison.  

In order to have a quantitative measure of the `variability' of the lightcurves, we estimate the standard deviation of the burst from an average profile that best imitates the burst profile. We construct a `trapezoidal' function with a rising part, a plateau and a tail for this purpose. The function is determined by five parameters: normalization $f0$ and nodal points $a, b, c$ and $d$. The rising part of the function is given as $f(t) = f0 (t-a)/(b-a)$, the plateau is the constant $f(t) = f_0$, and the tail is $f(t) = f0(d-x)/(d-c)$. After each run, we fitted the entire output SSC lightcurve with the trapezoidal function by varying the five parameters mentioned above, and estimated the standard deviation from the best fit trapezoid. We compare this value with the standard deviation obtained for the optical lightcurve, which is obtained by following the same method.

For a given distribution of $\Gamma_{300}$ and $R_{16}$ across the pulses, the best SSC amplification occurs when $\nu_c < \nu_{\rm{V}} < \nu_m$ and $\nu_c^{\rm{IC}} < \nu_m^{\rm{IC}} < \nu_\gamma$. The spectral regime $\nu_\gamma < \nu_c^{\rm{IC}} < \nu_m^{\rm{IC}}$ can also produce similar amplification, but for $\nu_c^{\rm{IC}}$ to be below $\sim 650$~keV one requires very small ($\sim 1$ - $10$~G) co-moving magnetic field. Only alternative scenarios involving quick decay of magnetic fields in the shock downstream (for example, \cite{2006ApJ...653..454P}) can achieve such low magnetic fields for standard input parameters. Moderate amplification can be obtained also if $\nu_c < \nu_m < \nu_{\rm{V}}$ and $\nu_c^{\rm{IC}} < \nu_\gamma < \nu_m^{\rm{IC}}$. But this spectral combination also requires the co-moving magnetic field to be too low. We could not find any condition where large variability amplification is possible. Hence we conclude that variabilities in the synchrotron lightcurve can be moderately amplified in the SSC lightcurve.

%
%
%
\begin{figure*}
\includegraphics[scale=0.40]{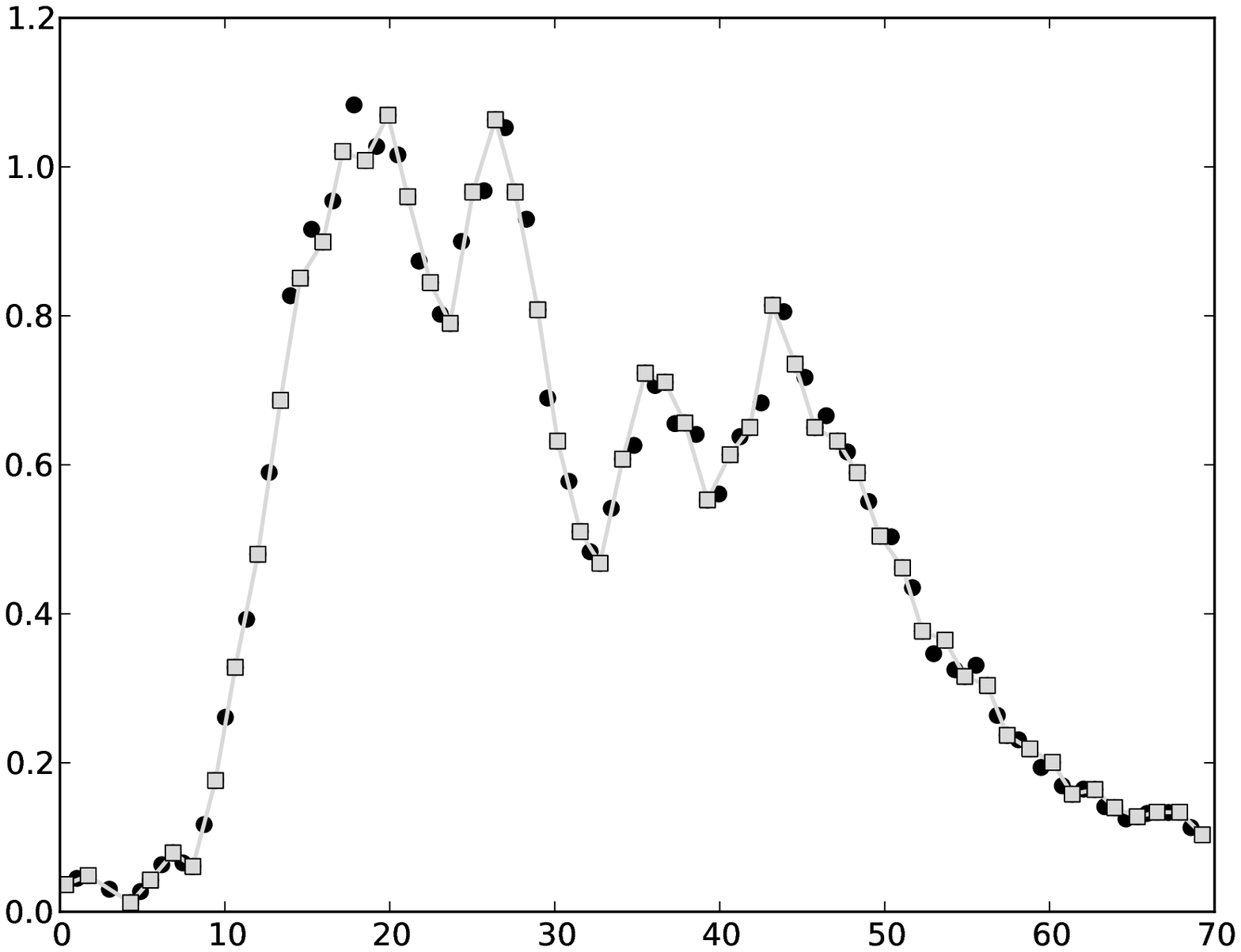}
\includegraphics[scale=0.40]{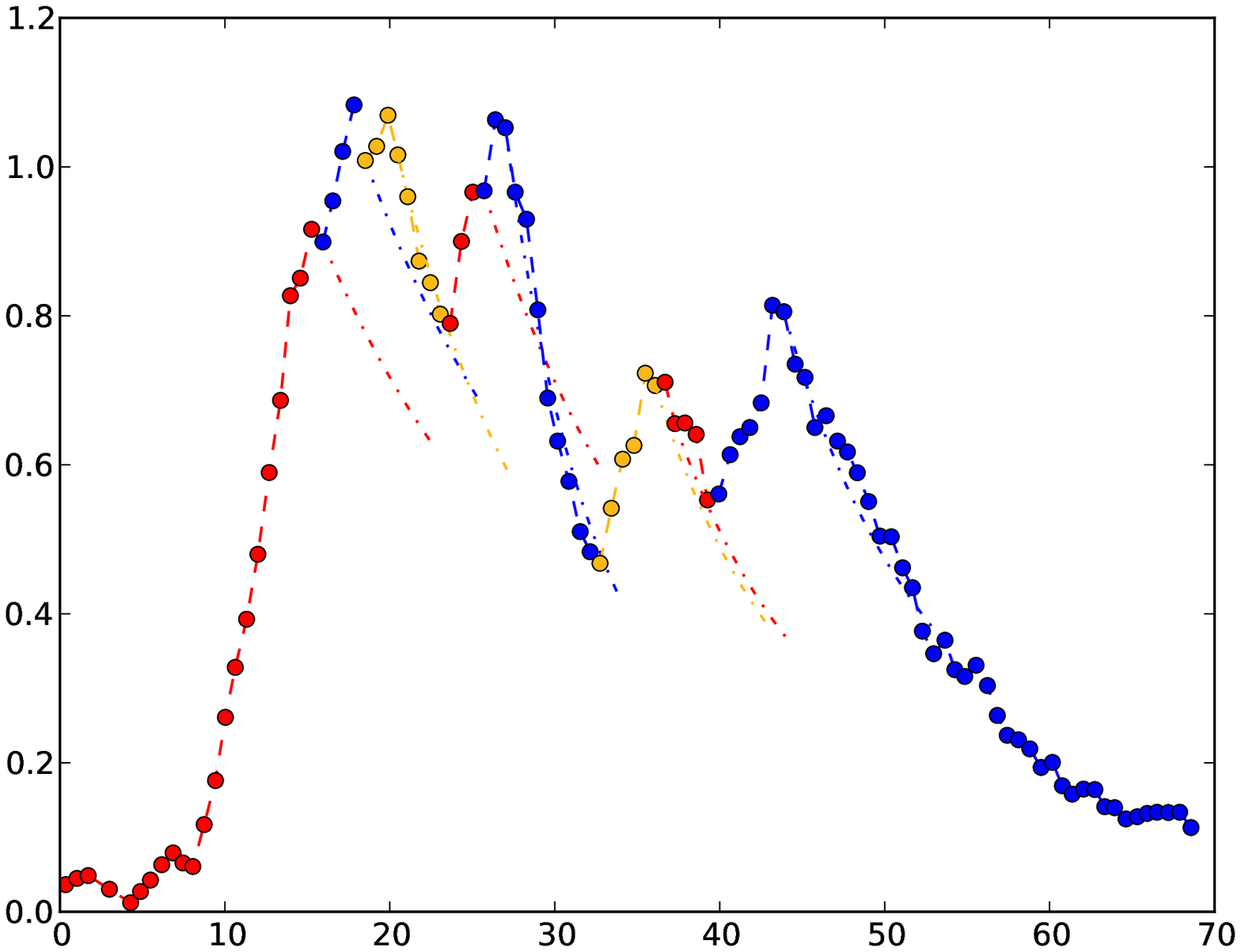}
\put(-385,0){Time since burst (sec)}
\put(-165,0){Time since burst (sec)}
\put(-465,40){\rotatebox{90}{$\frac{L_{\nu_V}}{10^{36}}$ erg$/$sec$/$Hz}}
\put(-225,40){\rotatebox{90}{$\frac{L_{\nu_V}}{10^{36}}$ erg$/$sec$/$Hz}}
\caption{The input synchrotron lightcurve. {\textsl{Left:}} The original data from Beskin et al. is shown in open squares connected by gray line. In black circles are the synthesized random fluctuations. {\textsl{Right :}} The eight pulses as we define them presented in different colors. The tails are computed using the curvature effect and are shown in dash-dotted line. See text for details.}
\end{figure*}
%
\begin{figure*}
\centerline{
\includegraphics[scale=0.5]{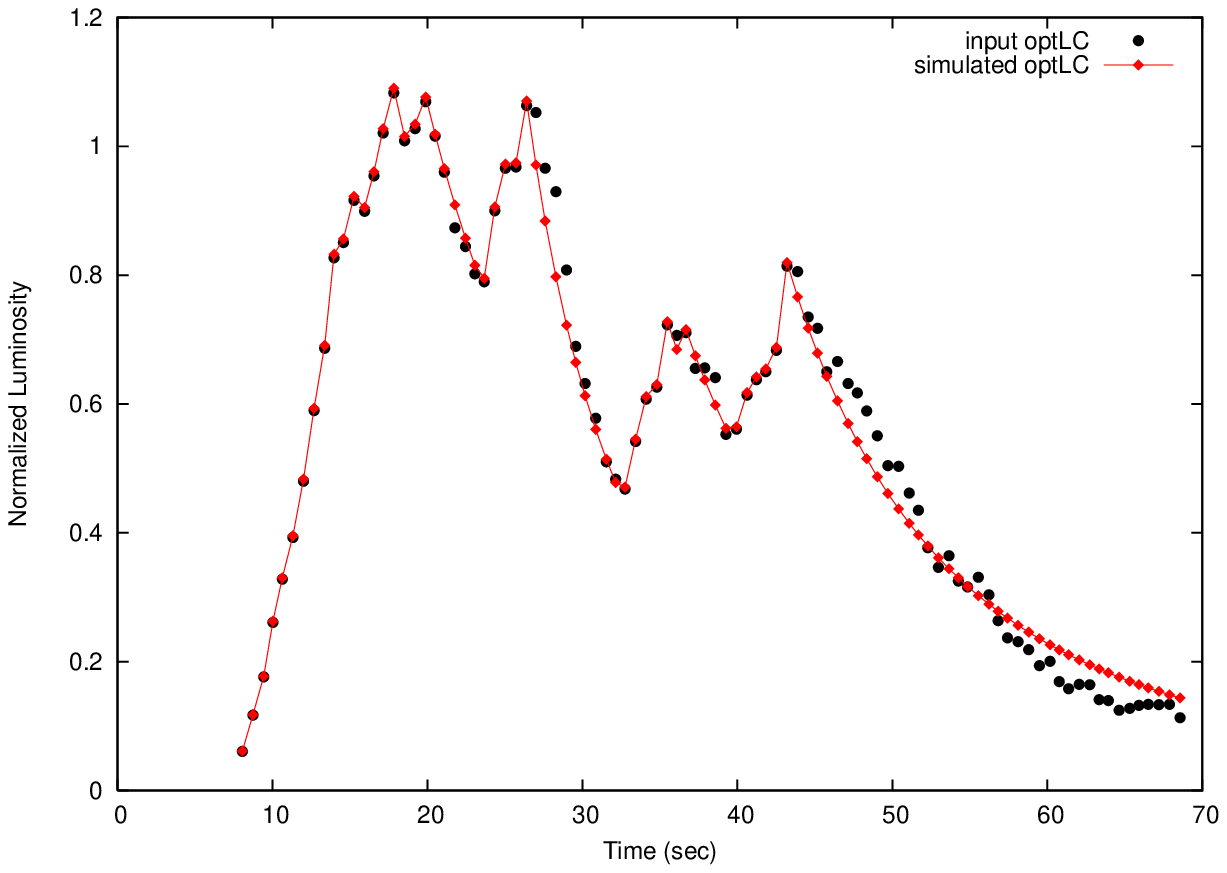}
\includegraphics[scale=0.50]{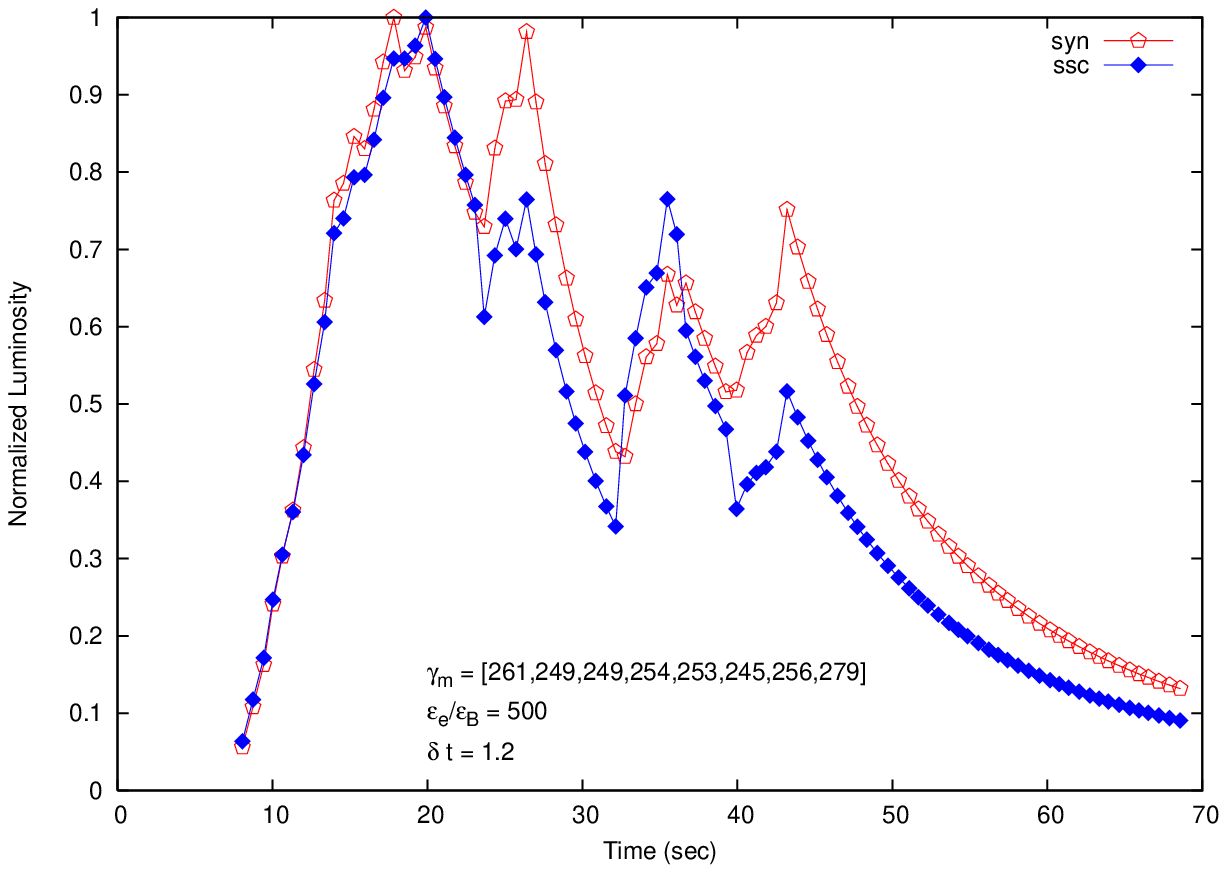}
}
\centerline{
\includegraphics[scale=0.500]{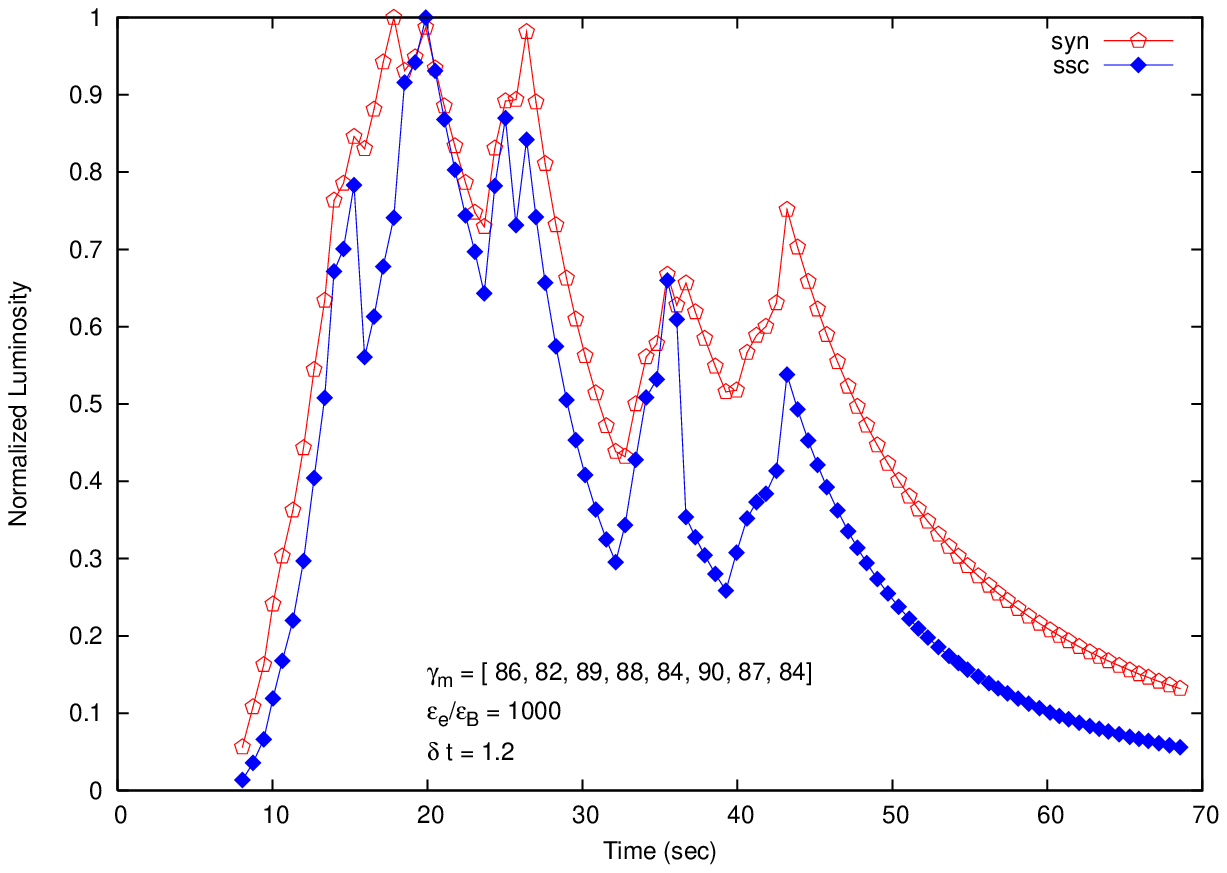}
\includegraphics[scale=0.500]{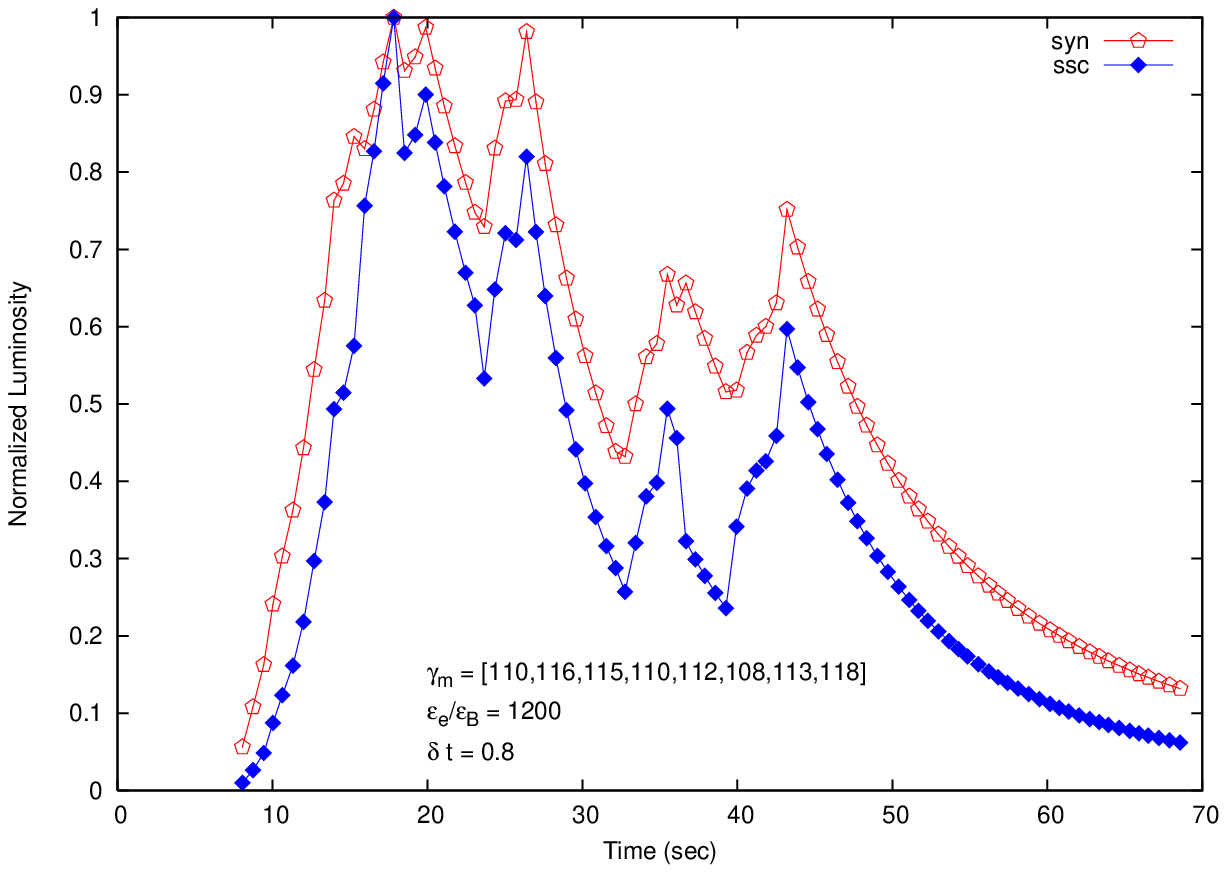}
}
\centerline{
\includegraphics[scale=0.500]{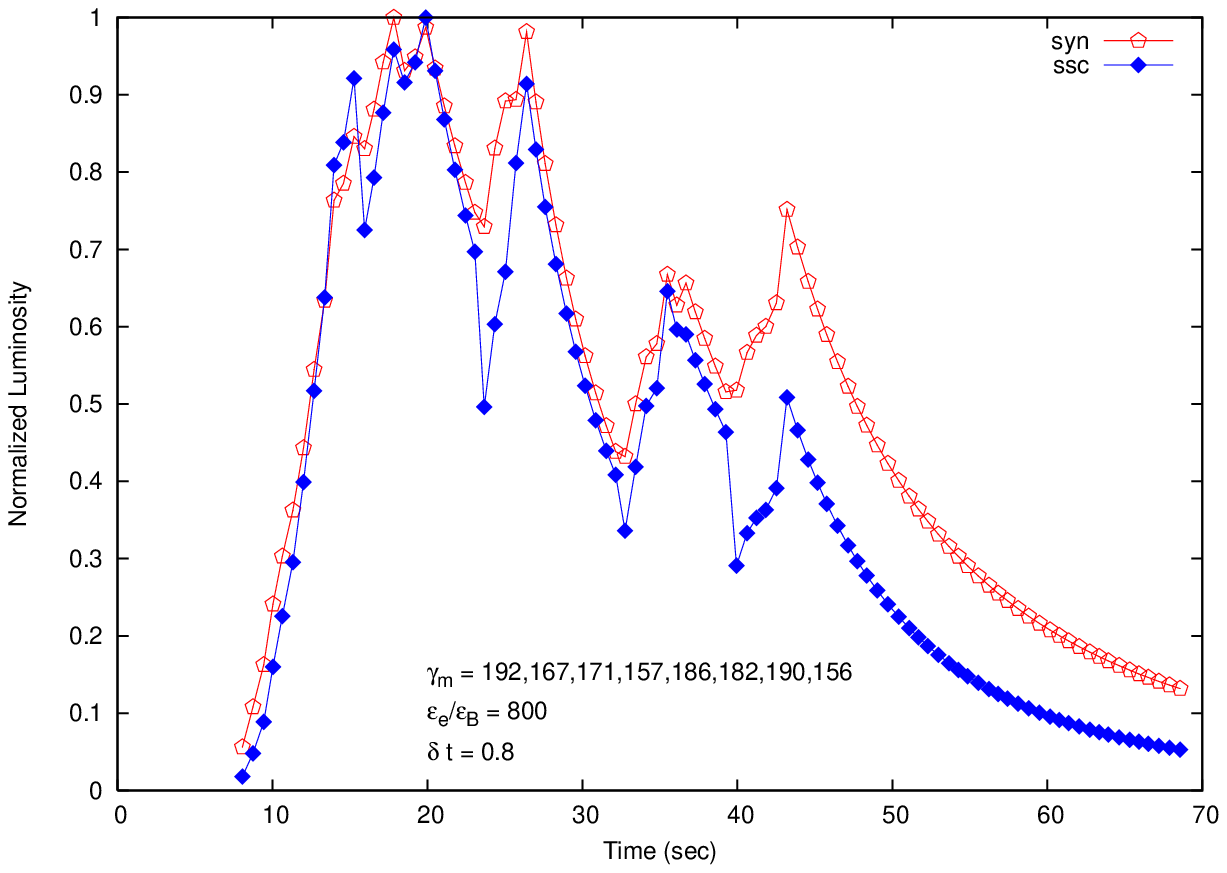}
\includegraphics[scale=0.500]{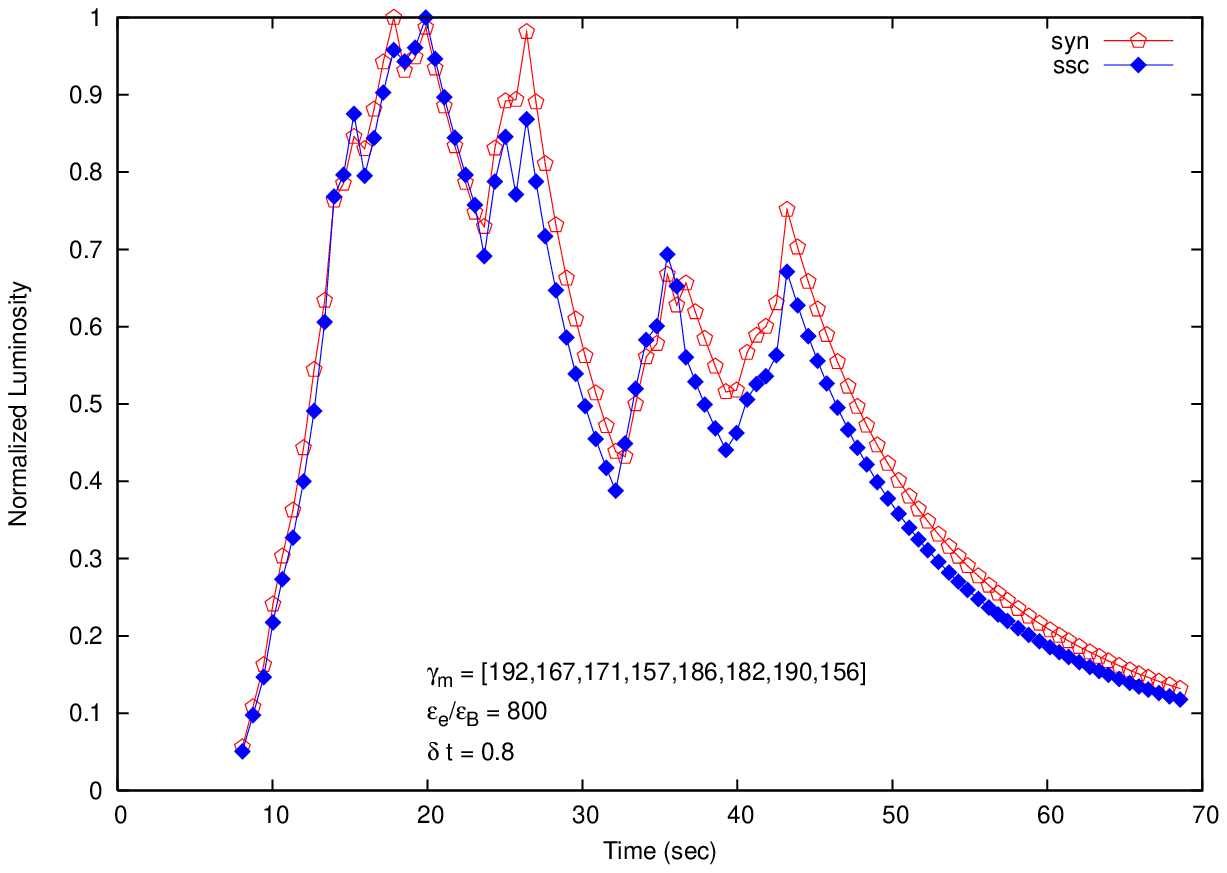}
}

\caption{In the first panel, we have plotted the input synthesized optical lightcurve (black) and the simulated lightcurve (red) from one of the runs. The rest of the panels (2 to 6), with normalized synchrotron luminosity in red and SSC in blue, showcase a variety of SSC lightcurves from our simulations. $\gamma_m$ for the individual pulses and the values of $\epsilon_e/\epsilon_B$ and $\delta t$ for each runs are included in the panels. Computed values of $\Gamma_{300}$ and $R_{16}$ for each runs are listed below. The numbers correspond to individual pulses. For panel 2 \& 3 : $\Gamma_{300} =\{0.759, 0.85, 0.787,0.8, 0.839, 0.717, 0.984, 0.86\},  R_{16} = \{ 0.124, 0.144, 0.134, 0.119, 0.122, 0.121, 0.18, 0.128\}$ , for panel 4 : $\Gamma_{300} =\{ 0.83, 0.778, 0.787, 0.801, 0.839, 0.716, 0.984, 0.859\},  R_{16} = \{9.9, 8, 8.9, 7.9, 8.2, 8.1, 12.0, 8.6 \} \times 10^{-2}$, for panel 5 \& 6 : $\Gamma_{300} =\{ 1.4, 1.42, 1.4, 1.47, 1.51, 1.37, 1.48, 1.5\},  R_{16} = \{0.42, 0.4, 0.42, 0.4, 0.4, 0.44, 0.39\}$. The only difference between panel 5 and panel 6 is the location of $\nu_{\gamma}$ in the SSC spectrum. In 5 $\nu_{\gamma} > \nu_m^{IC}$ while in 6 it is the opposite. In all panels, $\nu_c < \nu_V < \nu_m$. Both $\nu_V$ and $\nu_{\gamma}$ being in the same region of the spectrum diminishes the variability.}
\end{figure*}

\section{The `bottom-up' approach and simulated lightcurves}
\label{btmup}

To understand the variability better and also to have a complete picture, we approach the problem from the opposite end. We simulate the burst lightcurves from the `bottom-up' method where the final flux is built up from the input physical parameters of the emitting plasma. We assume the internal shock framework where the burst lightcurve is the sum of several independent pulses produced from random collisions between shells. We use a toy model where the temporal structure of a single pulse is determined by the evolution of the wind luminosity alone. However, more detailed modelling of the internal shock dynamics, including hydrodynamic simulations have been done in past \citep{2000A&A...358.1157D}.

We consider an isotropic wind luminosity $L_w(t)$ which varies between $t_0$ and $t_p$ as a shallow (as $t^{\delta}$ with $\delta$ between $0.5 - 1.0$) function of the observed time $t$ for a given pulse (ie., for one collision). $t_0$ is the time of collision and $t_p$ is the shock crossing time (also the pulse peak). The total number of radiating electrons $N_{e}$ is $\int_{0}^{t} L_w(t^{\prime}) dt^{\prime}/(\Gamma m_p c^2)$. In normalized units,
\be
N_{\rm{tot}, 55} = \frac{L_{w,52} t}{450 (\delta+1) \Gamma_{300}}
\ee

The co-moving magnetic field ($B^{\prime}$) is calculated by assuming that a fraction $\epsilon_B$ of the shock created thermal energy ($\theta_P m_p c^2$) will be carried by the magnetic field, where $\theta_P$ is the internal lorentz factor of the shocked shell. The downstream magnetic field density $u_{B^{\prime}} = \frac{N_p}{V^{\prime}} \theta_P m_p c^2$, where $N_p$, the number of protons ejected $\sim N_e$, and the co-moving volume $V^{\prime} = 4 \pi r^2 \Delta r$, with the thickness $\Delta r$ approximated as $c t \Gamma_{\rm{sh}}$. This leads to

\be
B^{\prime} = 473 \sqrt{\frac{L_{w,52} \theta_P \epsilon_B}{(\delta+1)R_{16}^2 \Gamma_{300}^2}}
\ee
The minimum lorentz factor $\gamma_m$ of the shock accelerated electron distribution is assumed to be 
\be
\gamma_m = \epsilon_e (m_p/m_e) \theta_P 
\ee
where $\epsilon_e$ is the fractional energy carried by these electrons. 

Total number of radiating electrons, $N_{\rm{rad}}$, is different from $N_{\rm{tot}}$ if the plasma is in `severe fast cooling', where time-scale ($t_{\rm{\gamma=1}}^{\prime}$) for electrons to cool down and loose all their kinetic energy is less than the dynamical time scale. In that case, a non-negligible fraction of electrons would pile-up at $\gamma$ of unity and would not be available in the relativistic pool to radiate via the non-thermal processes. This fraction keeps on increasing as the source ages. An exact estimate requires solving the continuity equation involving electron injection and radiative losses, which is beyond the scope of this paper. When $t_{\rm{\gamma=1}} \ll t_{\rm{dyn}}^{\prime}$ (`severe fast cooling'), the fraction of electrons remaining in the power-law can be assumed to be $\frac{t_{\rm{\gamma=1}}^{\prime}}{t_{\rm{dyn}}^{\prime}}$. Hence, the total number of radiating electrons, $N_{\rm{rad}} = \phi_{\rm{PL}} N_{\rm{tot}}$ where,
\be
\phi_{\rm{PL}} = {\rm{Min}} \Lb 1, \; \; \frac{t_{\rm{\gamma=1}}^{\prime}}{t_{\rm{dyn}}^{\prime}} \Rb,
\ee
$t_{\rm{\gamma=1}}^{\prime}$ is the time scale measure in the co-moving frame for an electron to cool down to $\gamma \sim 1$, which can be roughly expressed as $\frac{6 \pi m_e c}{\sigma_T} \, \frac{1}{(1+\mathcal{Y}_{\rm{Th}}+\mathcal{Y}_{\rm{Th}} \mathcal{Y}_{\rm{KN}}) {B^{\prime}}^2}$ (the derivation is given in the appendix).
It is relevant now to derive the expression for the cooling break in the electron spectrum $\gamma_c$. $\gamma_c \rightarrow 1$ in the `severe fast cooling' regime. During the normal fast cooling, it can be expressed as $\frac{t_{\rm{\gamma=1}}^{\prime}}{t_{\rm{dyn}}^{\prime}}$, which is same as $\phi_{\rm{PL}}$. Hence we have the following :

$
\gamma_c =
\begin{cases}
 \phi_{\rm{PL}} & \mbox {if    } t_{\rm{\gamma=1}} \gg  t_{\rm{dyn}}^{\prime} \\
 1   & \mbox {if    }  t_{\rm{\gamma=1}} \ll  t_{\rm{dyn}}^{\prime} 
\end{cases}
$

$
\frac{N_{\rm{rad}}}{N_{\rm{tot}}} =
\begin{cases}
 1 & \mbox {if    } t_{\rm{\gamma=1}} \gg  t_{\rm{dyn}}^{\prime} \\
 \phi_{\rm{PL}} &  \mbox {if    }  t_{\rm{\gamma=1}} \ll  t_{\rm{dyn}}^{\prime}
\end{cases}
$

The Compton $\mathcal{Y}$ parameter for the first order SSC scattering is calculated as $10^{-6} N_{\rm{rad},52} \gamma_m \gamma_c /R_{16}^2$ (eq-\ref{eqtau}) since $N_{\rm{rad}}$ and $B^{\prime}$ are known. $\mathcal{Y}$ parameter for the 2nd order SSC scattering is estimated by eq-\ref{yknexp}. 

We calculate the lightcurves in both slow and fast cooling cases, unlike in the previous approach. The synchrotron and SSC spectral breaks and peak flux are now calculated following the standard procedure. The synchrotron and SSC spectrum are calculated as piece-wise powerlaws, as described in section-2. After the pulse peak (shock crossing time), we calculate the flux decay using the curvature effect.

\begin{figure*}
\includegraphics[scale=0.4]{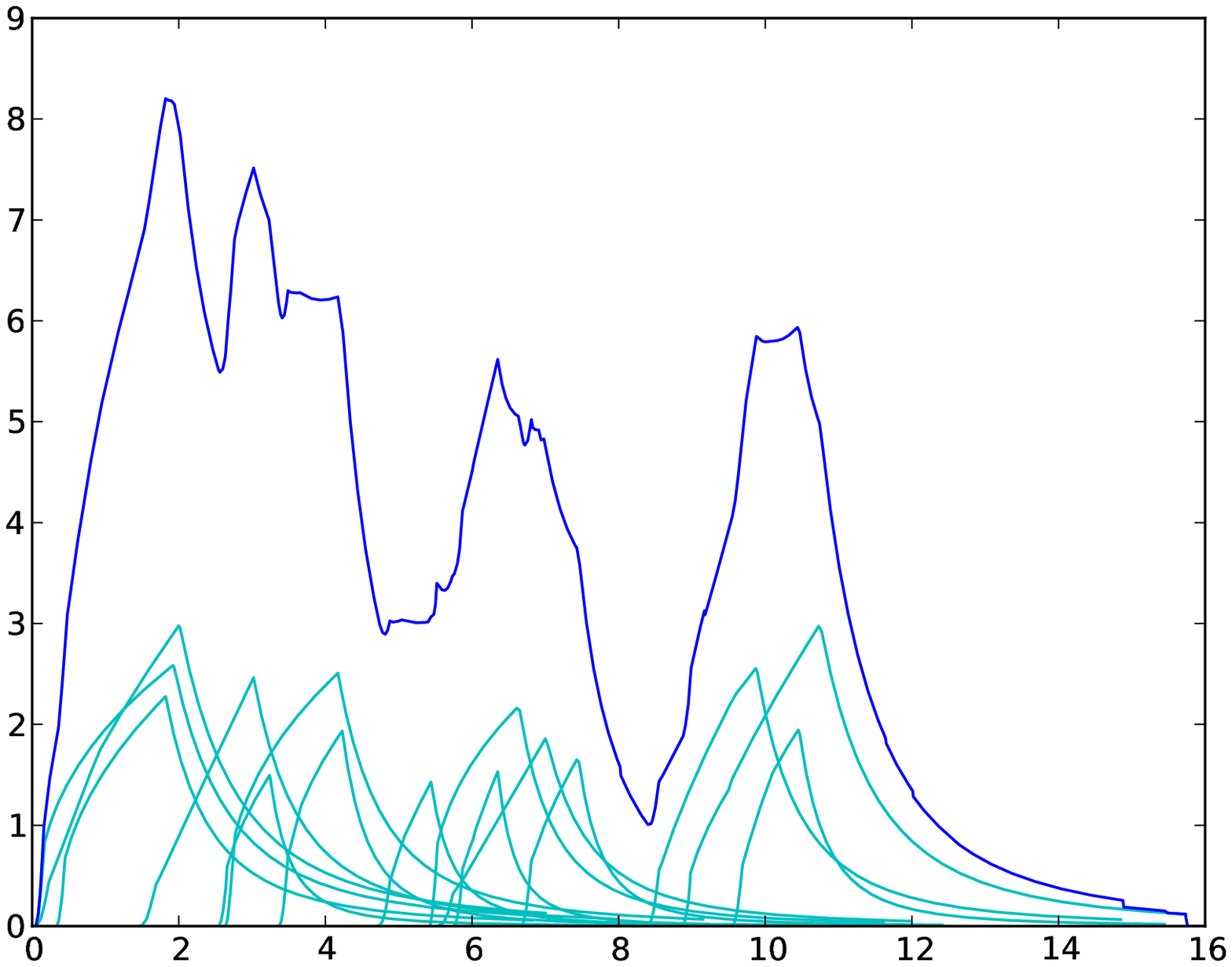}
\includegraphics[scale=0.4]{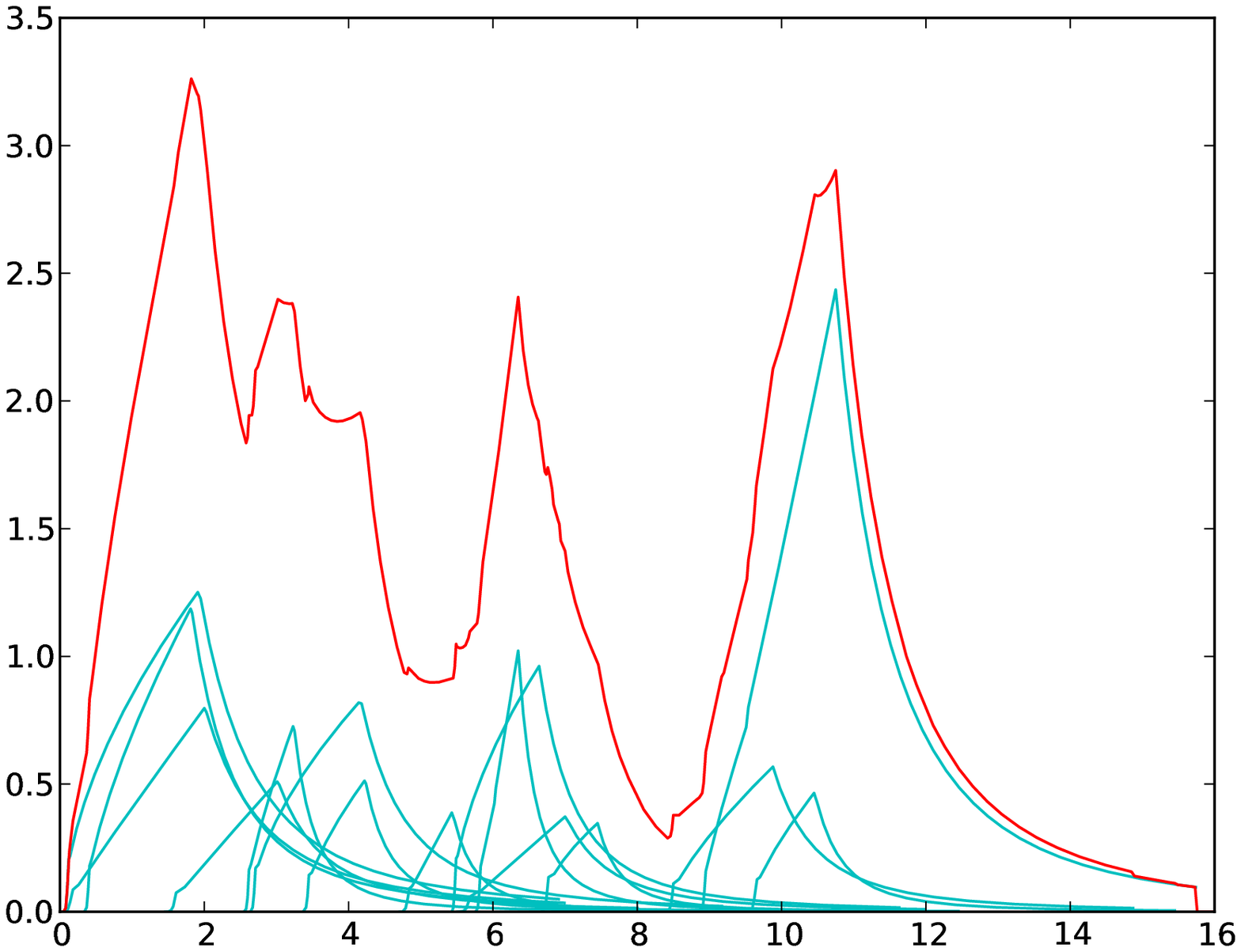}
\put(-392,2){Time since burst (sec)}
\put(-165,2){Time since burst (sec)}
\put(-450,40){\rotatebox{90}{Normalized Luminosity}}
\put(-220,40){\rotatebox{90}{Normalized Luminosity}}
\label{burstfig}
\caption{Typical synchrotron (left) and SSC (right) burst lightcurves from the simulations. The individual pulses are also shown. The typical time resolution we have is $\sim 0.01$~sec in the observed frame, hence that is limit of the sharpest features from the simulation. In this simulation, we have used $\epsilon_e = 0.01, \epsilon_B = 10^{-4}$ and $p=2.2$. $\Gamma_{300}$ was varied from $1.0$ to $2.5$, $R_{16}$ from $0.08$ to $0.1$, $\theta_p$ from $3.3$ to $5.5$, and $\delta$ from $0.5$ to $0.9$. $L_{w,52}$ was kept at $15 (t/{\rm{1 sec}})^{\delta}$.}
\end{figure*}

We ran a Monte-carlo simulation where the luminosity normalization, temporal index $\delta$ and $\theta_P$ are considered as random variables with a uniform distribution. The luminosity normalization and $\theta_P$ range between the typical values of $1.0 - 15.0$ and $3.0 - 10.0$ respectively. Either from observations or theory, there is not a clear idea about the values $\delta$. A time independent wind luminosity can not produce a rising, fast-cooling synchrotron pulse, hence we chose a shallow range of $0.5 - 1.0$ for $\delta$. We also consider the shock crossing time (or pulse peak) to be distributed randomly between $0.5 - 2.0$. Each pulse thus generated are later shifted by a random $t_0$ value. The final burst profile is the sum of all these individual pulses (see figure-\ref{burstfig}). Like in the `top-down' formalism, we scan the typical ranges of the parameters. $R_{16}$ is scan from $0.01$ to $1.0$ and $\Gamma_{300}$ from $0.3$ to $3$. In a single simulation, the radius and $\Gamma$ are changed by a small factor. A range of $0.01 - 0.1$ in $\epsilon_e$ and a range of $10^{-4} - 5 \times 10^{-4}$ is scanned in $\epsilon_B$. In a single run, these parameters are kept fixed. Value of $p$ is fixed at $2.2$. Even though the typical temporal index for the rising phase of an individual pulse is $3\delta/4$, if the observed frequency is between $\nu_c$ and $\nu_m$ much sharper observed rise profiles are possible due to the cumulative effect of multiple pulses.

\subsection{Measure of variability} 
We estimate the `variability' following the same method as in the `top-down' scenario. We obtain the standard deviation of both synchrotron and SSC lightcurves from the best-fit trapezoid. We find that the SSC lightcurves are at best only slightly more variable than the synchrotron lightcurve. With some set of parameters, the SSC variability can even be less than the synchrotron one. This is because the overall flux variation in the SSC is much lesser than its synchrotron counter part. The number of extra `spikes', even if present in the SSC, do not contribute much in the estimated variability.
 
The luminosity normalization and $\theta_P$ have more influence in the amplification of the pulses compared to $\delta$. We made runs by keeping $\delta$ as a constant and also by allowing it to vary. Low variation in both $L_w$ and $\theta_P$ and constant $\delta$ can result in nearly identical pulse shapes between synchrotron and SSC components. We did not find any marked difference between results from slow and fast cooling cases.
%

%
\section{Summary and discussion}
\label{Conclusions}
In this paper, we have done a comparative study of lightcurve variability of the synchrotron and SSC components of GRB prompt emission. Starting from a template synchrotron lightcurve with small scale fluctuations, we trace back the magnetic field of the emitting region and self-consistently calculate the SSC lightcurve using assumptions for the electron distribution. We investigate how the small scale fluctuations in the SSC lightcurve are related to the initial variabilities present in the synchrotron lightcurve. A multitude of temporal structures can be obtained for the SSC lightcurve depending on the parameters of the electron distribution function. Degree of modification of the input fluctuations are different for different spectral regimes within the SSC spectrum, due to varying sensitivity to $\gamma_m$. Miniscule changes in the electron distribution, which appear as indiscernible pulses in the synchrotron lightcurve, can get moderately amplified in the SSC lightcurve. Hence, in general, the SSC lightcurve can be more variable compared to the synchrotron lightcurve, however, not to a large extent. We complement the formalism with a `bottom-up' approach where the synchrotron and SSC lightcurves are calculated through a Monte-Carlo simulation of internal shock model. Only moderate amplification to the variabilities could be obtained in this approach as well. 

We apply our method to the ``naked-eye'' GRB 080319B. This burst has been interpreted within the framework of the synchrotron + SSC model \citep{2008MNRAS.391L..19K,2008Natur.455..183R}. Several difficulties have been raised for this model, including the time lag between the $\gamma$-ray and optical lightcurves, as well as the energy budget crisis from the second-order SSC. Here we apply another criterion, the relative variability between the two components, to investigate the validity of the model. 
 We create a synchrotron template based on the observed optical data of GRB 080319B (with minor modifications to test how small variabilities are amplified), and calculate the expected $\gamma$-ray variabilities in various spectral regimes. We found that the model lightcurves are all smoother than the observed one. We then conclude that the optical/$\gamma$-ray lightcurves are difficult to account for within the simplest synchrotron/SSC model. 

This conclusion is reached based on our analytical formalism presented in this paper. For an analytical treatment to be possible, we have to adopt some approximations and/or make some assumptions. We conclude by listing them and commenting on their validity. 
\begin{itemize}
\item The total bolometric luminosity is a fraction of the power dissipated by the internal shocks, which in turn is a fraction of the wind luminosity from the central engine. Such a treatment has been adopted in all internal shock model calculations in the past.
\item The ratio $\epsilon_e/\epsilon_B$, indicating the fractional energy in electrons and in the magnetic field respectively, is assumed to be a constant. Numerical simulations have started to derive $\epsilon_e$ and $\epsilon_B$ from the first principle, but none have revealed how these parameters depend on shock parameters. In principle, both values may evolve with shock parameters. However, since little understanding is achieved regarding such an evolution, we take the simplest assumption that the {\em ratio} $\epsilon_e/\epsilon_B$ is constant. We note that in afterglow modeling, the assumption of a constant $\epsilon_e$ and $\epsilon_B$ throughout the deceleration phase seems to fit the observed data well.
\item The bulk Lorentz factor $\Gamma$ and the emission radius $R$ are allowed to randomly vary from pulse to pulse. By doing so, we have assumed that multiple collisions happen at different radii between shells of varying lorentz factor. This variation did not significantly affect the relative variability contrast between the synchrotron and SSC lightcurves.
\item In both `top-down' and `bottom-up' models, we have not considered the evolution of $R$ and $\Gamma$ in one collision.
\item In the `top-down' method, the underlying electron distribution is approximated to be in the fast-cooling regime, i.e. even the cooling time of the electrons with the lowest injection energy is shorter than the dynamical time scale. Such an approximation is found self-consistent given the typical value of $B^{\prime}$ we derive. In the `bottom-up' method, both slow and fast cooling conditions are tested.
\item The power-law index of the injected spectrum is assumed to be $2.2$, a typical value for relativistic shocks. Our results do not significantly depend on this value. Changing it to a different value would not affect our conclusion.
\item The synchrotron spectrum is approximated as a multi-segment broken power-law representing different spectral regimes. The analytical treatment of SSC \citep{2007MNRAS.380...78G} is adopted. In the expression for the peak luminosity of the synchrotron spectrum, we have neglected a correction factor that takes care of the contribution from the power-law electron distribution \citep{1999ApJ...523..177W, 2007MNRAS.380...78G}. The SSC peak luminosity is assumed to be $\mathcal{Y}_{\rm{Th}}$ times the synchrotron peak. All these are standard analytical treatments of synchrotron and SSC emission spectra. More realistic treatments would not affect the variability contrast in the synchrotron and SSC lightcurves.
\item For the second order scattering in the Klein-Nishina regime, the $\mathcal{Y}$-parameter is approximately estimated by scaling down its value for the first order scattering in the Thomson regime by a factor invoking the ratio between the KN cross section and the Thomson cross section at the electron injection energy.
\item The assumption of a fast cooling spectrum leads to the definition of the bolometric synchrotron luminosity $L_{\rm{syn}} = L_{\nu_m} \nu_m$. An accurate treatment would introduce a correction factor of the order of unity, but would not affect our conclusion.
\end{itemize}
We thank the anonymous referees for constructive comments that improved the quality of the paper. This work is supported by NASA NNX09AO94G, NNX10AD48G and NSF AST-0908362. LR acknowledges support from the French Agence Nationale de la Recherche via contract ANR-JC05-44822.
%

%
%
\section*{Appendix : Detailed Expressions}

For $\nu_c < \nu_V < \nu_m$, 
\be
L_{\rm{bol},52} \simeq 0.26 \, {L_{V,36}}^{4/3} \Lb \frac{\nu_V}{5.45 \times 10^{14}} \Rb^{2/3} \Lb \frac{a_g^2-1}{a_g} \Rb^{1/3} \frac{\epsilon_e}{\epsilon_B} \Lb \frac{\gamma_m}{200}\frac{1}{\sqrt{R_{16}} \mathcal{Y}_{\rm{Th}}} \Rb^{4/3}
\label{maineq-1}
\ee
and
\be
B^{\prime} \simeq 100 {\rm{G}} \, {L_{V,36}}^{2/3}  \Lb \frac{\nu_V}{5.45 \times 10^{14}} \Rb^{1/3} \Lb \frac{a_g^2-1}{a_g} \Rb^{2/3} \frac{1}{\Gamma_{300}}  \Lb \frac{\gamma_m}{200}\frac{1}{R_{16}^2 \mathcal{Y}_{\rm{Th}}} \Rb^{2/3}
\label{maineq-2}
\ee

For $\nu_c < \nu_V <\nu_m$, 
\be
L_{\rm{bol},52} \simeq 0.08 \, {L_{V,36}}^{20/21} \Lb \frac{\nu_V}{5.45 \times 10^{14}} \Rb^{22/21} \Lb \frac{a_g^2-1}{a_g} \Rb^{1/3} \frac{\epsilon_e}{\epsilon_B} \frac{R_{16}^{2/21}}{ {\mathcal{Y}_{\rm{Th}}}^{20/21}} \Lb \frac{500}{\gamma_m} \Rb^{4/21}
\label{maineq-3}
\ee
and 
\be
B^{\prime} \simeq 53 {\rm{G}} \, {L_{V,36}}^{10/21} \Lb \frac{\nu_V}{5.45 \times 10^{14}} \Rb^{11/21} \Lb \frac{a_g^2-1}{a_g} \Rb^{2/3} \frac{1}{\Gamma_{300} R_{16}^{20/21} {\mathcal{Y}_{\rm{Th}}}^{10/21}} \Lb \frac{500}{\gamma_m} \Rb^{2/21}
\label{maineq-4}
\ee

After substituting for $\gamma_{\rm{KN},2}$ as $500 {B^{\prime}}^{-1/5}$ (see section-2 about how we obtained this expression) we can rewrite eq-\ref{yknexp} as 

\be
\mathcal{Y}_{\rm{KN}} = 0.3 {\Lb\frac{B^{\prime}}{300} \Rb}^{-6/5} {\Lb\frac{\gamma_m}{200}\Rb}^{-6} \mathcal{Y}_{\rm{Th}} \, \frac{3}{8x_{\gamma_m}} \Lb \log{2x_{\gamma_m}}+\frac{1}{2} \Rb
\ee

where $x_{\gamma_m}$ defined in section--- in the same normalized units is $ \sim 5 \frac{B^{\prime}}{300} {\Lb \frac{\gamma_m}{200} \Rb}^{5}$.

Hence the equality eq-\ref{root} can be written as 

\be
\mathcal{Y}_{\rm{Th}} (1 + \mathcal{Y}_{\rm{Th}} + \mathcal{A}(\gamma_m, B^{\prime}) {\mathcal{Y}_{\rm{Th}}}^2  ) = \frac{\epsilon_e}{\epsilon_B} 
\label{root2}
\ee

where the term $ 0.3 {\Lb\frac{B^{\prime}}{300} \Rb}^{-6/5} {\Lb\frac{\gamma_m}{200}\Rb}^{-6} \, \frac{3}{8x_{\gamma_m}} \Lb \log{2x_{\gamma_m}}+\frac{1}{2} \Rb $ which essentially is the product $\cal R_{\sigma} \times \Delta \mathcal{E}_\mathcal{R}$ is written as $\mathcal{A}(\gamma_m, B^{\prime})$. In case if the 2nd order IC scattering is in the KN-regime (which can happen for high values of $\gamma_m$), $\mathcal{A}$ will be unity as both $\cal R_{\sigma}$ and $\Delta \mathcal{E}_\mathcal{R}$ will be reduced to unity.

\section*{Appendix : Calculation of electrons in the `cooling pile'}
If the radiative cooling is severe, a fraction of electrons cool down, loose all their kinetic energy within the dynamical timescale and pile up around $\gamma = 1$. We estimate the remaining fraction in the powerlaw as $\frac{t_{\gamma=1}}{t_{\rm{age}}} \times N_{\rm{tot}}$, where $t_{\gamma=1}$ is the time scale for an electron to cool down to $\gamma = 1$. $t_{\gamma=1}$ can be estimated from the cooling rate as follows:

The radiative cooling is due to both synchrotron and IC processes. The total energy loss rate $\frac{d \gamma}{dt}$ can be written as $\left({\frac{d \gamma}{dt}}\right)_{\rm{syn}} + \left({\frac{d \gamma}{dt}}\right)_{\rm IC}$. The second term include two components, i.e. energy loss due to $1^{\rm{st}}$ and $2^{\rm{nd}}$ order IC, which can be approximated as $\mathcal{Y} \left( \frac{d \gamma}{dt}\right)_{\rm{syn}}$ and $\mathcal{Y} \mathcal{Y}_{\rm{KN}} \left(\frac{d \gamma}{dt}\right)_{\rm{syn}}$, respectively. The synchrotron loss itself can be approximated as $\left(\frac{d \gamma}{dt}\right)_{\rm{syn}} = \frac{\sigma_T}{6 \pi m_e c} \frac{1}{\gamma^2}$. As a result, the total energy loss of an electron can be written as
\be
\frac{d \gamma}{dt} = 1.3 \times 10^{-9} {B^{\prime}}^2 \gamma^2 (1+\mathcal{Y}_{\rm{Th}}+\mathcal{Y}_{\rm{Th}} \mathcal{Y}_{\rm{KN}})~, 
\label{eqdiff}
\ee
where the constant $1.3 \times 10^{-9}$ is from $\frac{\sigma_T}{6 \pi m_e c}$.

Ideally $\mathcal{Y}_{\rm{KN}}$ is a function of the electron lorentz factor $\gamma$ and magnetic field is a function of time. In our treatment, we are focused on an average estimate. We use the expression derived for $\mathcal{Y}_{\rm{Th}}$ and $\mathcal{Y}_{\rm{KN}}$ in Sect. \ref{sec23}, and hence take out the $\gamma$ dependence of the coefficients in eq-\ref{eqdiff}. We use the magnetic field at the peak of the pulse as a representative value. We can then solve the differential equation under the assumption that the magnetic field is a constant over time, which is not a bad assumption as long as the cooling time scale is much shorter than the dynamical time scale. 
\be
\gamma (t) = \frac{\gamma_0}{1+ 1.3 \times 10^{-9} (1+\mathcal{Y}_{\rm{Th}}+\mathcal{Y}_{\rm{Th}} \mathcal{Y}_{\rm{KN}}) \gamma_0 {B^{\prime}}^2 t}
\ee

Under heavy radiative loss, the term unity in the denominator can be neglected, so that one has  $\gamma(t) \approx  \frac{1}{1.3 \times 10^{-9} (1+\mathcal{Y}_{\rm{Th}}+\mathcal{Y}_{\rm{Th}} \mathcal{Y}_{\rm{KN}}) {B^{\prime}}^2 t}$. Substituting for $\gamma (t) = 1$, one can then obtain the approximate $t_{\gamma =1} = {\LB 1.3 \times 10^{-9} (1+\mathcal{Y}_{\rm{Th}}+\mathcal{Y}_{\rm{Th}} \mathcal{Y}_{\rm{KN}}) \, {B^{\prime}}^2\RB}^{-1}$.
\end{document}